\documentclass[%
 reprint,
 amsmath,amssymb,
 aps,
]{revtex4-2}

\usepackage{graphicx}
\usepackage{dcolumn}
\usepackage{bm}
\usepackage{float} 
\usepackage{booktabs} 
\usepackage{makecell} 
\usepackage{lineno}
\usepackage{comment}

\begin{document}

\preprint{APS/123-QED}

\title{Amplitude Analysis of the Invariant-Mass Spectra of Centrally Exclusively Produced \(\pi^+\pi^-\) and \(K^+K^-\) Pairs in \(pp\) Collisions at \(\sqrt{s}=510\) GeV }

\author{Aranya Giri}
\email{aranyagiri230597@gmail.com}
\affiliation{%
Department of Physics, University of Houston\\
Houston, Texas - 77204
}

\author{Barak Schmookler}
\email{baschmookler@uh.edu}
\affiliation{%
Department of Physics, University of Houston\\
Houston, Texas - 77204
}

\author{Rene Bellwied}
\email{bellwied@uh.edu}
\affiliation{%
Department of Physics, University of Houston\\
Houston, Texas - 77204
}



\begin{abstract}
We present an amplitude analysis of the $\pi^{+}\pi^{-}$ and $K^{+}K^{-}$ invariant-mass spectra from central exclusive $pp$ collisions at $\sqrt{s}=510~\mathrm{GeV}$, measured by the STAR experiment with forward-scattered protons tagged in the Roman Pot detectors. The $\pi^{+}\pi^{-}$ channel is described with $f_{0}(500)$, $\rho^{0}(770)$, $f_{0}(980)$, $f_{2}(1270)$, $f_{0}(1500)$ and $f_{0}(1710)$ resonances with $\chi^{2}/\mathrm{ndf}=121.712/57$. The fit also yields $m_{f_{0}(980)}=0.989$ GeV and $\Gamma=0.091$ GeV. The $K^{+}K^{-}$ channel is described with $f_{0}(980)$, $f_{2}(1270)$, $f_{2}^{\prime}(1525)$ and $f_{0}(1710)$ resonances with $\chi^{2}/\mathrm{ndf}=14.662/15$. The fiducial cross sections for the individual resonances and interference terms were calculated, together with their ratios relative to the total fitted cross section. These ratios were compared across decay channels and qualitatively with previous measurements. A consistent flavor dependence of $f_{0}(1500)$ and $f_{0}(1710)$ has been derived for the highest RHIC collision energies, which is in agreement with low energy data.
\end{abstract}

\maketitle



\section{\label{sec:level1} Introduction}

\begin{figure}[H]
    \centering
    \includegraphics[width=0.75\linewidth]{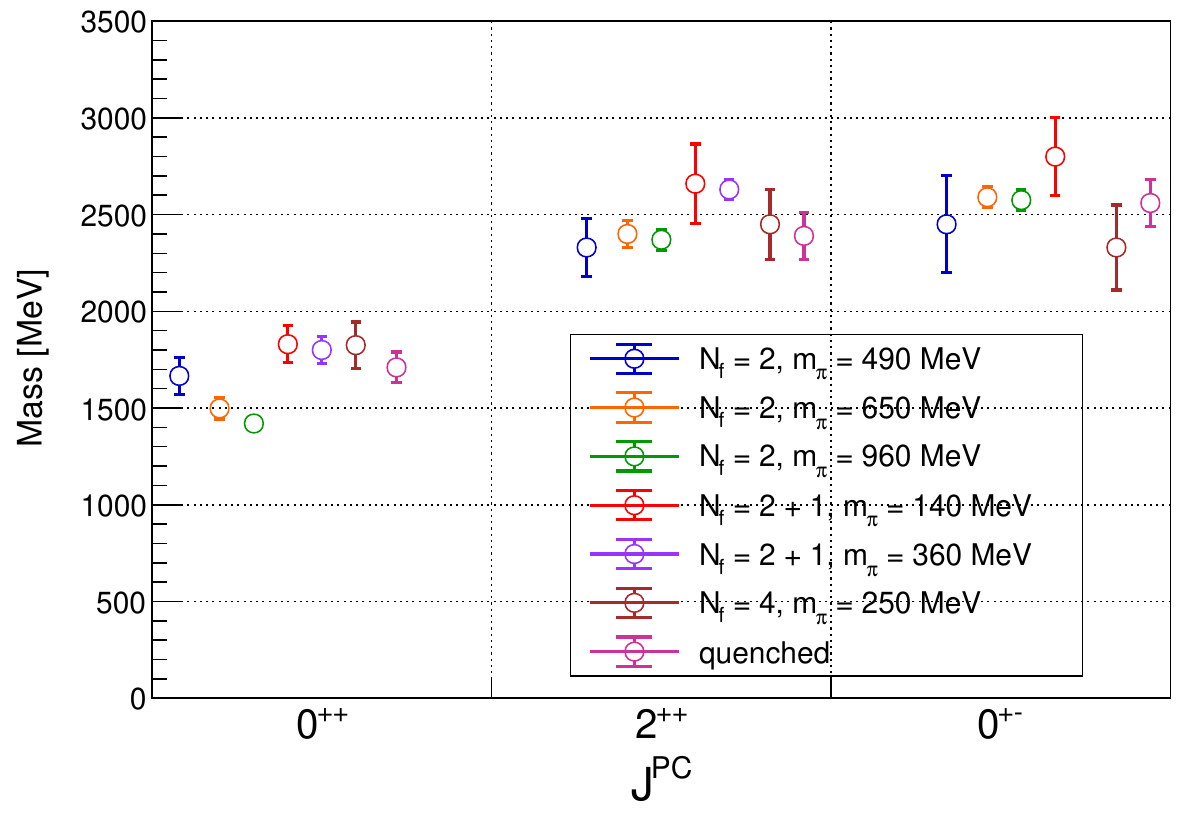}
    \caption{Lattice-QCD predictions for the glueball spectrum. Figure adapted from Ref.~\cite{Vadacchino2023}.}
    \label{fig:lattice_predictions}
\end{figure}

Quantum chromodynamics (QCD) predicts the existence of color-singlet bound states composed entirely, or predominantly, of gluons, known as \emph{glueballs}.
Unlike conventional mesons, which are formed from quark--antiquark ($q\bar{q}$) pairs, glueballs arise from the self-interaction of gluons, a unique consequence
of the non-Abelian nature of QCD. Lattice-QCD (LQCD) calculations consistently predict that the lightest glueball possesses scalar quantum numbers, $J^{PC}=0^{++}$, with a mass in the range of $1.4$--$1.8~\mathrm{GeV}/c^{2}$~\cite{Vadacchino2023}. The LQCD predictions for the low-lying glueball spectrum are
shown in Fig.~\ref{fig:lattice_predictions}.

Among the experimentally observed isoscalar scalar mesons, the $f_{0}(1500)$ and $f_{0}(1710)$ have long been regarded as the leading candidates for scalar-glueball. However, the identification of the scalar glueball is complicated by the fact that a glueball carries the same quantum numbers as conventional
isoscalar $q\bar{q}$ mesons. Consequently, an observed physical state may be a pure glueball, a pure quarkonium, or mixed state containing both gluonic and
$q\bar{q}$ components. Determining the exact nature of these glueball candidates remains one of the central challenges in hadron spectroscopy.

Central exclusive production (CEP) provides one of the cleanest experimental environments for hadron spectroscopy. In CEP, the centrally produced hadronic
system is separated from the outgoing beam particles by large rapidity gaps, while both incident protons remain intact,
\begin{equation}
    \label{cep_equation}
    p + p \rightarrow p' + X + p',
\end{equation}
where $X$ denotes the centrally produced hadronic system and $+$ denotes the rapidity gaps.

\begin{figure}[!h]
    \centering
    \includegraphics[width=\linewidth]{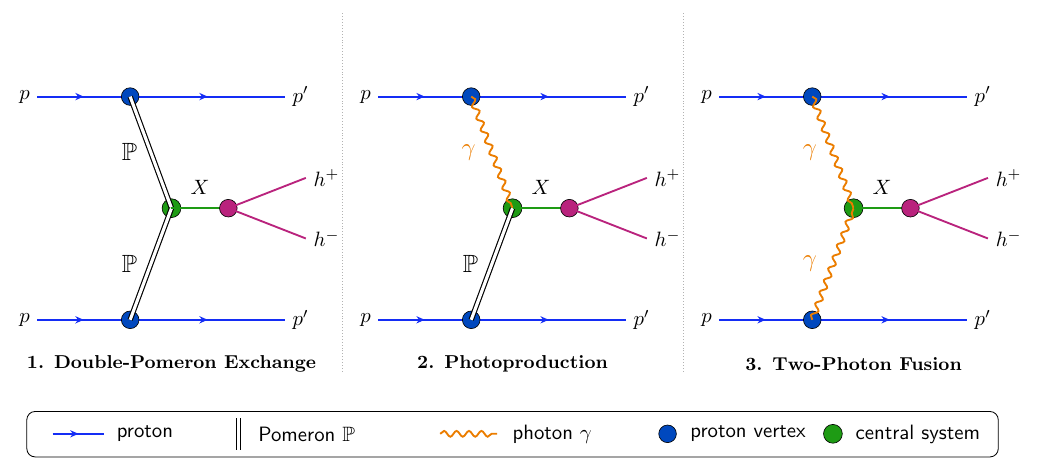}
    \caption{Mechanisms contributing to central exclusive production.}
    \label{fig:cep}
\end{figure}

\begin{table*}[t]
\caption{Summary of resonances observed by WA102 in different decay channels.}
\label{tab:WA102}
\centering
\small
\setlength{\tabcolsep}{6pt}
\renewcommand{\arraystretch}{1.15}

\begin{ruledtabular}
\begin{tabular}{lccc}
Decay channel & $S$ wave & $P$ wave & $D$ wave \\
\hline
$\pi^+\pi^-$ ~\cite{WA102_pi+pi-}
& $f_0(980)$, $f_0(1370)$, $f_0(1500)$, $f_0(1710)$
& $\rho(770)$
& $f_2(1270)$ \\

$\pi^0\pi^0$ ~\cite{WA102_pi0pi0}
& $f_0(980)$, $f_0(1370)$, $f_0(1500)$
& ---
& $f_2(1270)$ \\

$K^+K^-$ ~\cite{WA102_KKbar}
& $f_0(980)$, $f_0(1500)$, $f_0(1710)$
& ---
& $f_2(1270)$, $f_2'(1525)$, $f_2(2150)$ \\

$K_S^0K_S^0$ ~\cite{WA102_KKbar}
& $f_0(980)$, $f_0(1500)$, $f_0(1710)$
& ---
& $f_2(1270)$, $f_2'(1525)$ \\

$\eta\eta$ ~\cite{WA102_EtaEta}
& $f_0(1370)$, $f_0(1500)$, $f_0(1710)$
& ---
& $f_2(1270)$, $f_2(2150)$ \\

$\pi^+\pi^-\pi^+\pi^-$ ~\cite{WA102_4pion}
& $f_0(1500)$, $f_0(2000)$
& $f_1(1285)$
& $f_2(1950)$, $\eta_2(1620)$, $\eta_2(1875)$

\end{tabular}
\end{ruledtabular}
\end{table*}

The central system can be produced through pomeron--pomeron ($\mathbb{P}\mathbb{P}$), photon--pomeron ($\gamma\mathbb{P}$), or photon--photon ($\gamma\gamma$)
interactions, as illustrated in Fig.~\ref{fig:cep}. For the production of low-mass hadronic states in proton--proton collisions at RHIC energies, the dominant
contribution arises from $\mathbb{P}\mathbb{P}$ interactions, while the $\gamma\mathbb{P}$ $\gg$ $\gamma\gamma$ processes are expected to be significantly low.
The Pomeron (\(\mathbb{P}\)) is commonly modeled as a predominantly gluonic, color-singlet exchange carrying the quantum numbers of the vacuum. Consequently,
conservation of quantum numbers in Pomeron–Pomeron (\(\mathbb{P}\mathbb{P}\)) interactions favors the production of isoscalar states with positive charge
conjugation and parity. The centrally produced states are therefore expected to predominantly have quantum numbers $ I^{G}J^{PC}=0^{+}(0,2,\ldots)^{++}$. The
combination of the gluon-rich nature of Pomeron exchange and the quantum-number selection rules associated with \(\mathbb{P}\mathbb{P}\) interactions makes CEP
a particularly promising experimental environment for searching for and studying glueball candidates, especially the lightest scalar glueball.

\subsection{Previous Experimental Observation in CEP}
\label{sec:previous_measurements_in_cep}

The first extensive search of glueball candidates in this process was carried out by the WA102 experiment, which studied CEP in fixed-target proton--proton collisions using a $450~\mathrm{GeV}/c$ proton beam. Partial-wave analyses (PWAs) were performed for the centrally produced $\pi^+\pi^-$, $\pi^0\pi^0$, $K^+K^-$, $K_S^0K_S^0$, $\eta\eta$, and $\pi^+\pi^-\pi^+\pi^-$ final states. The resonances observed in the $S$-, $P$-, and $D$-wave amplitudes across the different decay channels are summarized in Table~\ref{tab:WA102}.

At the LHC, the CMS Collaboration measured $\pi^+\pi^-$ CEP in $pp$ collisions at $\sqrt{s}=5.02$ and $13~\mathrm{TeV}$~\cite{CMS2020}. However, because the
scattered protons were not detected, the selected event sample was not fully \textit{exclusive} and could contain contributions from proton dissociation. A complex amplitude model incorporating the nonresonant continuum and the $f_{0}(500)$, $\rho^{0}(770)$, $f_{0}(980)$, and $f_{2}(1270)$ resonances was employed to describe the dipion invariant-mass spectrum. No significant contributions from the $f_{0}(1500)$ and $f_{0}(1710)$ resonances were reported. It is important to emphasize that the sensitivity of the CMS measurement to resonant structures in the $m_{\pi\pi}\gtrsim 1.5~\mathrm{GeV}$ region was limited by the substantial non-exclusive multiparticle background, as illustrated in Fig.~4 of Ref.~\cite{CMS2020}, and the resulting deterioration of the signal-to-background ratio. Consequently, the uncertainty associated with the multiparticle-background subtraction increased substantially above $m_{\pi\pi}=1.5~\mathrm{GeV}$, ranging from approximately $20\%$ to $60\%$. This reduced sensitivity may have hindered the reliable identification of potential scalar glueball candidates in this mass region. Nevertheless, the amplitude-based fit to the dipion invariant-mass spectrum reported by CMS is adopted in our analysis, as it yields a $\rho^{0}$ production cross section that is slightly higher than, but comparable to, the STARLight Monte Carlo (MC) prediction. 

At RHIC, the STAR Collaboration measured CEP in \(pp\) collisions at \(\sqrt{s}=200~\mathrm{GeV}\) and \(510~\mathrm{GeV}\)~\cite{STAR_CEP_200GeV,STARCEP_510GeV}. At \(\sqrt{s}=200~\mathrm{GeV}\), STAR measured differential cross sections and, after extrapolation to a Lorentz-invariant kinematic region, described the \(\pi^+\pi^-\) invariant-mass spectrum using a phenomenological model consisting of a nonresonant continuum together with contributions from the \(f_0(980)\), \(f_2(1270)\), and \(f_0(1500)\) resonances (the fit model is similar to CMS). This procedure involves model-dependent extrapolations of the produced-pion and scattered-proton kinematics into phase-space regions inaccessible to the detector, as discussed in Sec.~11.2 of Ref.~\cite{STAR_CEP_200GeV}. At \(\sqrt{s}=510~\mathrm{GeV}\), STAR compared the measured invariant-mass spectra with predictions from the GRANIITTI v1.0 CEP Monte Carlo event generator, which incorporates both nonresonant and resonant components. The analysis reports contributions from \(f_0(980)\), \(f_2(1270)\), and \(f_0(1710)\) in the channel \(\pi^+\pi^-\), and from \(f_0(980)\), \(f_0(1500)\), \(f'_2(1525)\), and \(f_0(1710)\) in the channel \(K^+K^-\). The limitations of the GRANIITTI MC comparison in assessing the sensitivity to glueball candidates, along with the motivation for reanalyzing the CEP data at \(\sqrt{s}=510~\mathrm{GeV}\), are discussed further in Section~\ref{sec:graniitti_mc_evgen}.

\subsection{Status of the $f_0(1500)$ and $f_0(1710)$ glueball candidates}
\label{sec:status_of_glueball}

A widely used phenomenological framework for interpreting these states is glueball--quarkonium mixing, in which the physical $f_0$ states are expressed as admixtures of a scalar glueball $G$, a strange configuration $s\bar{s}$, and a non-strange configuration $n\bar{n}\equiv(u\bar{u}+d\bar{d})/\sqrt{2}$. The inferred compositions are strongly model dependent, since they depend on the assumed masses of the unmixed (bare) states, the structure and strength of the mixing interactions, and the treatment of decay and production amplitudes. Close, Kirk and Zhao (CKZ) modeling favor the bare-state ordering $M_{s\bar{s}}>M_{G}>M_{n\bar{n}}$. The model found a substantial glueball component in the $f_0(1500)$, and $s\bar{s}$ in the $f_0(1710)$~\cite{WA102_CEP_mixing,radiative_jpsi_mixing}. This assignment is consistent with quenched LQCD calculations available at that time, which predicted a glueball mass of approximately $1.55$ GeV~\cite{QuenchedLatticeGlueball_Old}, and is also consistent with the recent Belle measurement of $\Gamma_{\gamma\gamma}\mathcal{B}(f_0(1710)\to K\bar{K})$ in the $\gamma\gamma\to K_{S}^{0}K_{S}^{0}$ channel. Belle obtained a relatively large value of $\Gamma_{\gamma\gamma}\mathcal{B}(K\bar{K}) =12^{+3+227}_{-2-8}\,\mathrm{eV}$, which is difficult to reconcile with the strongly suppressed two-photon coupling expected for a pure glueball, $\Gamma_{\gamma\gamma}\ll 1\,\mathrm{eV}$. This photon coupling result favors $f_0(1710)$ having a substantial $s\bar{s}$ component~\cite{Belle_KSKS_2013} a  the CKZ model predicts.

In contrast, the modeling of Cheng, Chua, and Liu (CCL) is motivated by the latest quenched LQCD results~\cite{QuenchedLatticeGlueball_New}. They fix the bare glueball mass to the LQCD value of $1.71$ GeV and assume an approximately flavor-$SU(3)$-symmetric scalar quarkonium spectrum, $M_{s\bar{s}}\simeq M_{n\bar{n}}$. This leads to a qualitatively different assignment: the $f_0(1710)$ is predominantly glueball, and the $f_0(1500)$ is dominated by $s\bar{s}$~\cite{ChengChuaLiu2006}. This assignment is supported by the production rates measured in radiative $J/\psi$ decays. The inferred production rate of the $f_0(1710)$, $\mathcal{B}(J/\psi\to\gamma f_0(1710))\sim 2.2\times10^{-3}$, is of the same order as the LQCD prediction for a scalar glueball, $\mathcal{B}(J/\psi\to\gamma G)\simeq 3.8(8)\times10^{-3}$. In contrast, the corresponding production rate of the $f_0(1500)$ is only $\sim 0.29\times10^{-3}$, approximately an order of magnitude smaller than that of the $f_0(1710)$. This difference in their production rates in the gluon-rich radiative $J/\psi$ decay environment is consistent with CCL modeling~\cite{JinShen2021}.

Despite these observations, no concrete interpretation of the scalar states in this mass region has yet emerged. This work presents an amplitude analysis of the previously published STAR central exclusive production $\pi^{+}\pi^{-}$ and $K^{+}K^{-}$ invariant-mass spectra at $\sqrt{s}=510~\mathrm{GeV}$~\cite{STARCEP_510GeV}. The primary objective is to investigate the contributions of the $f_{0}(1500)$ and $f_{0}(1710)$, two prominent scalar-glueball candidates in the $1.5$--$1.7~\mathrm{GeV}$ mass region, and to evaluate their impact on the description of the measured spectra. Their relative contributions to the $\pi^{+}\pi^{-}$ and $K^{+}K^{-}$ final states provide complementary information on their decay patterns, enabling comparison with previous measurements and offering further input for their interpretation as glueball candidates.

\section{Sensitivity of \textsc{GRANIITTI} to the \(f_0(1500)\) and \(f_0(1710)\) Signals}
\label{sec:graniitti_mc_evgen}

\textsc{GRANIITTI} is a Monte Carlo event generator developed for high-energy diffractive scattering, with particular emphasis on central exclusive and semi-exclusive processes of the form
\begin{equation}
    pp \rightarrow p^{(*)} + X + p^{(*)},
    \label{eq:cep-process}
\end{equation}
where \(X\) denotes a centrally produced system and \(p^{(*)}\) represents either an intact proton or a low-mass proton-dissociative state~\cite{Graniitti}. One of the principal mechanisms implemented in \textsc{GRANIITTI} is double-Pomeron exchange (DPE). The spin of the Pomeron is modeled as \(J=0\) for scalar and tensor resonances and as \(J=1\) for pseudoscalar resonances~\cite{Graniitti}. \textsc{GRANIITTI} also includes photon-induced processes, with the photon flux calculated using the equivalent-photon approximation, including its transverse-momentum dependence~\cite{Graniitti}. For two-meson final states, the model includes both a nonresonant continuum and the production and decay of intermediate resonances. Schematically, the coherent mass-dependent amplitude may be represented as
\begin{equation}
    \mathcal{M}(m)
    =
    \mathcal{M}_{\mathrm{cont}}(m)
    +
    \sum_R c_R e^{i\phi_R}\mathcal{M}_R(m),
    \label{eq:graniitti-amplitude}
\end{equation}
where \(\mathcal{M}_{\mathrm{cont}}\) denotes the continuum amplitude, \(\mathcal{M}_R\) is the mass-dependent amplitude for resonance \(R\), \(c_R\) is its relative magnitude, and \(\phi_R\) is its relative phase. The observable spectrum is proportional to \(\lvert\mathcal{M}\rvert^2\).

The performance of \textsc{GRANIITTI} v1.0 can be assessed through comparisons with the STAR measurements of
central exclusive meson-pair production at $\sqrt{s}=200$ and $510~\mathrm{GeV}$. The model parameters are tuned
using the $200~\mathrm{GeV}$ data~\cite{STARCEP_510GeV}. Figure~4 of Ref.~\cite{Graniitti_FitTo200GeV} compares \textsc{GRANIITTI} calculations with the STAR measurements of exclusive $\pi^{+}\pi^{-}$ and $K^{+}K^{-}$ production at $\sqrt{s}=200~\mathrm{GeV}$. For the $\pi^{+}\pi^{-}$ channel, \textsc{GRANIITTI} reproduces the overall shape of the invariant-mass spectrum and describes the dominant structures associated with the $f_{0}(980)$ and $f_{2}(1270)$ resonances. However, noticeable deviations remain, particularly above $M_{\pi\pi}\approx1.5~\mathrm{GeV}$, where additional scalar and tensor resonances, including the $f_{0}(1500)$ and $f_{0}(1710)$, contribute. Similarly, for the $K^{+}K^{-}$ channel, the model reproduces the general trend of the measured spectrum, although the central prediction deviates from the data over a significant fraction of the invariant-mass range. While the relatively large experimental uncertainties partially obscure these discrepancies in the $K^{+}K^{-}$ channel, the overall agreement remains primarily qualitative. 

The \textsc{GRANIITTI} v1.0 predictions were also compared with the independent STAR CEP measurements at $\sqrt{s}=510~\mathrm{GeV}$~\cite{STARCEP_510GeV}. The $510~\mathrm{GeV}$ measurement was performed within a slightly different kinematic acceptance from that used at $200~\mathrm{GeV}$. As shown in Fig.~8 of Ref.~\cite{STARCEP_510GeV}, the model does not provide a quantitative description of the measured invariant-mass spectra without additional normalization adjustments. Specifically, normalization factors of 0.49 and 1.0 were applied to the resonance-plus-continuum predictions for the $\pi^{+}\pi^{-}$ spectra with $\Delta\varphi<90^{\circ}$ and $\Delta\varphi>90^{\circ}$, respectively, while factors of 1.45 and 2.16 were applied to the corresponding $K^{+}K^{-}$ spectra. The need for these channel- and kinematics-dependent normalization factors limits the extent to which the comparison can be used to assess the contributions of individual resonances, particularly those of the $f_{0}(1500)$ and $f_{0}(1710)$.
 
Overall, these comparisons demonstrate that the GRANIITTI v1.0 implementation provides a useful phenomenological framework for studying CEP, incorporating both nonresonant and resonant contributions and qualitatively reproducing the main features of the measured spectra. At the same time, the discrepancies observed at both \(\sqrt{s}=200\) and \(510~\mathrm{GeV}\), indicate that the present model does not yet provide a complete quantitative description of the data. In particular, the resonance content of the higher invariant-mass region, \(m\gtrsim1.5~\mathrm{GeV}\), remains insufficiently constrained. This motivates a complementary, fit-driven investigation in which individual resonance amplitudes can be systematically introduced or removed and their contributions assessed through changes in the fit quality. Following the amplitude-fitting approach employed by CMS~\cite{CMS2020} and STAR at \(\sqrt{s}=200~\mathrm{GeV}\)~\cite{STAR_CEP_200GeV}, the present analysis examines the possible contributions of the \(f_{0}(1500)\) and \(f_{0}(1710)\) resonances by incorporating their amplitudes individually and in combination. The resulting changes in the $\chi^2$ provide a means of assessing the evidence for these states and comparing their contributions in the \(\pi^{+}\pi^{-}\) and \(K^{+}K^{-}\) decay channels.


\section{ Brief Details of the Dataset}
\label{sec:dataset_details}

The analysis uses the STAR central exclusive production (CEP) data collected in proton--proton collisions at $\sqrt{s}=510$ GeV during the 2017 RHIC run, corresponding to an effective integrated luminosity of $25.6\,\mathrm{pb}^{-1}$ after correcting for the CEP trigger efficiency. Approximately $6.22\times10^{8}$ events satisfying the CEP trigger were recorded. The trigger required signals in the Roman Pot detectors on both sides of the interaction point, at least two hits in the Time-of-Flight (TOF) detector, and vetoes on the Beam--Beam Counters (BBC) and Zero Degree Calorimeters (ZDC) to ensure the characteristic double-rapidity-gap topology and suppress proton-dissociation and pile-up backgrounds.

During the offline event selection, events were required to contain exactly two forward proton tracks reconstructed within the Roman Pot fiducial acceptance and exactly two oppositely charged primary tracks reconstructed in the Time Projection Chamber (TPC) and matched to TOF hits. The charged-particle pairs were identified as $\pi^{+}\pi^{-}$, $K^{+}K^{-}$, or $p\bar{p}$ using the combined TPC specific ionization ($dE/dx$) and TOF information. Details of the Roman Pot fiducial acceptance, track-quality requirements, and particle-identification criteria can be found in Ref.~\cite{STARCEP_510GeV}.

The exclusivity of the selected events was ensured by requiring the missing transverse momentum of the complete final state,
\begin{equation}
 \left|\bm{p}_{T,p'p'}+\bm{p}_{T,h^+h^-}\right|<0.12~\mathrm{GeV}/c,
 \label{eq:ptmiss}
\end{equation}
which efficiently suppresses non-exclusive backgrounds. The remaining background contribution was estimated from the missing-transverse-momentum distribution and statistically subtracted from the invariant-mass spectra.\\

After applying all selection criteria, the final sample consisted of 86\,008 $\pi^{+}\pi^{-}$ candidates, 2\,454 $K^{+}K^{-}$ candidates, and 225 $p\bar{p}$ candidates. The estimated exclusive fractions are $80.6\%$ for the $\pi^{+}\pi^{-}$ sample, $87.9\%$ for the $K^{+}K^{-}$ sample, and $57.0\%$ for the $p\bar{p}$ sample. Since the sample $p\bar{p}$ has limited statistics, the present analysis focuses exclusively on the final states $\pi^{+}\pi^{-}$ and $K^{+}K^{-}$ Ref.~\cite{STARCEP_510GeV}. The $\pi^{+}\pi^{-}$ and $K^{+}K^{-}$ invariant-mass distributions from the two complementary scattered-proton azimuthal regions ($\Delta\phi < 90^{\circ}$ and $\Delta\phi > 90^{\circ}$) were combined to improve the statistical precision, particularly in the $K^{+}K^{-}$ channel, where the available statistics are more limited. The same approach was applied to the $\pi^{+}\pi^{-}$ channel to ensure a consistent treatment and facilitate direct comparisons between the two decay channels. Although the azimuthal regions were originally separated to investigate the proposed ``glueball-filter'' effect~\cite{CloseKirk1997}, this distinction is not essential to the present analysis, which focuses
on the decay-channel dependence of the resonance contributions rather than their dependence on the azimuthal separation of the scattered protons.


\section{Fit Model}
\label{sec:fit_model}

The measured differential cross section is modeled as the convolution
of the underlying invariant-mass spectrum with the detector
mass-resolution function:

\begin{align} 
\frac{d\sigma}{dm}&=\int dm'\,G(m,m';\sigma_m)\,[I_{\rm VMP}(m')+I_{\rm DPE}(m')],\\ 
I_{\rm VMP}&=\left|\sum_{R'}A^{\mathrm{VMP}}_{R'}\right|^2,\qquad I_{\rm DPE}=\left|\sum_R A^{\mathrm{DPE}}_R+A_{\mathrm{cont}}\right|^2. \label{eq:model} 
\end{align}

Here, \(G(m,m';\sigma_m)\) denotes the detector mass-resolution function. The mass dependence of the resolution, \(\sigma_m(m)\), and its parameterization are described in Appendix~\ref{app:mass_resolution}.

The two intensity terms in Eq.~\ref{eq:model} correspond to vector-meson photoproduction (VMP) and double-Pomeron exchange (DPE). The VMP and DPE contributions are added incoherently, since the interference between the two production mechanisms is expected to be small; an effect at the level of approximately \(1\%\)~\cite{CMS2020}. Within each production mechanism, however, the contributing amplitudes are summed coherently.

The nonresonant CEP contribution is included in the DPE amplitude and is parameterized as
\begin{equation}
A_{\mathrm{cont}}(m)
=
a_{\mathrm{cont}}\,
A_{\mathrm{cont}}^{\mathrm{PDF}}(m),
\label{eq:continuum_amplitude}
\end{equation}
where \(a_{\mathrm{cont}}\) is a free normalization parameter and \(A_{\mathrm{cont}}^{\mathrm{PDF}}(m)\) describes the invariant-mass dependence of the non-resonant continuum amplitude obtained from \textsc{GRANIITTI}~\cite{Graniitti}. For each decay channel, the simulated non-resonant continuum events are subjected to the same event-selection criteria as those applied in data~\cite{STARCEP_510GeV}.

For a resonance \(R\), the corresponding amplitude is parameterized as
\begin{equation}
A_R(m)
=
a_R e^{i\phi_R}\,
\mathrm{BW}_R(m),
\label{eq:resonance_amplitude}
\end{equation}
where \(a_R\) and \(\phi_R\) denote the magnitude and phase of the resonance amplitude, respectively. A relativistic Breit--Wigner line shape, defined in Appendix~\ref{app:breit_wigner}, is used for each resonance except the \(f_0(980)\), which is described using the Flatt\'e parameterization given in Appendix~\ref{app:flatte}.

The fit is performed using \textsc{Minuit2} within the ROOT framework. The resonance masses and widths are fixed to the values reported by the Particle Data Group~\cite{PDG2024}. But the mass and width of \(f_0(500)\) are fixed to the values reported in~\cite{GarciaMartin2011}, while the parameters that govern the shape of the \(f_0(980)\) Flatt\'e line are allowed to vary in fit. It should be emphasized that to reliably extract the resonance amplitudes and phases, parameters such as the resonance masses and widths should be kept fixed whenever possible. Allowing an excessive number of parameters to vary can lead to unstable fits and substantial fluctuations in the extracted amplitudes and phases.

The coherent \(f_{0}\)--\(f_{2}\) interference is retained in the present analysis because the orthogonality of the corresponding \(S\)- and \(D\)-wave spherical harmonics, which causes their interference to vanish upon integration over the full solid angle, is not necessarily preserved within a finite detector acceptance. To investigate this effect, a simple toy model (see Appendix~\ref{app:toy_orthogonality}) was constructed in which the centrally produced \(\pi^{+}\pi^{-}\) system contains both \(S\)- and \(D\)-wave components. In particular, WA102 observed the \(f_{2}(1270)\) and \(f_{2}^{\prime}(1525)\) predominantly in the \(D_{0}\) wave~\cite{WA102_pi+pi-,WA102_KKbar}, motivating the choice of the \(D_{0}\) component in the toy-model study. The normalized \(Y_{00}\)--\(Y_{20}\) overlap remains close to zero for nearly full angular coverage, consistent with the expected orthogonality, but becomes increasingly negative as the pion pseudorapidity acceptance is restricted. 

\begin{figure}[H]
    \centering
    \includegraphics[width=0.45\textwidth]{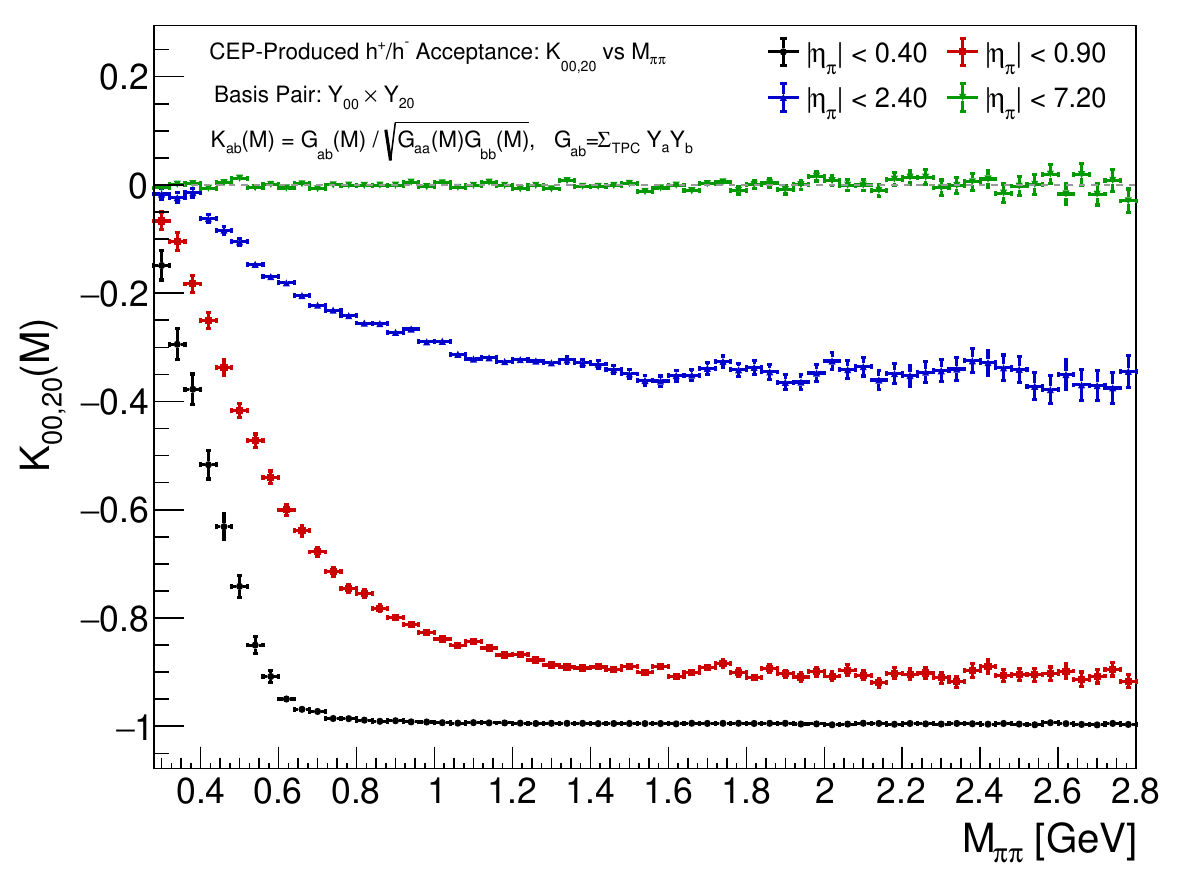}
    \caption{Normalized overlap \(K_{00,20}(M_{\pi\pi})\) between the \(Y_{00}\) (\(S\)-wave) and \(Y_{20}\) (\(D_{0}\)-wave) spherical harmonics as a function of the \(\pi^{+}\pi^{-}\) invariant mass for different pion pseudorapidity acceptances. The departure of \(K_{00,20}\) from zero with decreasing acceptance illustrates the breaking of the effective \(S\)--\(D_{0}\) orthogonality.}
    \label{fig:toy_model_Y00_Y20}
\end{figure}

As shown in Fig.~\ref{fig:toy_model_Y00_Y20}, in the higher invariant-mass region relevant to the \(f_{2}(1270)\) and \(f_{2}^{\prime}(1525)\), the normalized overlap \(K_{00,20}(M_{\pi\pi})\) reaches approximately \(-0.35\) for the CMS acceptance, \(|\eta_{\pi}|<2.4\)~\cite{CMS2020}, and \(-0.90\) for the STAR \(\sqrt{s}=510~\mathrm{GeV}\) acceptance, \(|\eta_{\pi}|<0.9\)~\cite{STARCEP_510GeV}. For still narrower acceptance, the overlap approaches \(-1\), indicating that the two angular components become nearly proportional with opposite signs within the accepted phase space. More generally, the magnitude of the overlap increases with \(M_{\pi\pi}\) before approaching an approximate plateau. These results demonstrate that restricted angular acceptance can break the effective orthogonality of the \(S\)- and \(D_{0}\)-wave components, allowing interference between coherent \(f_{0}\) and \(f_{2}\) amplitudes to survive angular integration. This acceptance effect therefore motivates retaining the \(f_{0}\)--\(f_{2}\) interference term in the present fit. Notably, an \(f_{0}\)--\(f_{2}\) interference term was also retained in the corresponding CMS and STAR \(\sqrt{s}=200~\mathrm{GeV}\) analyses, although the role of finite angular acceptance was not explicitly discussed.

\section{Fit Results}

\subsection{$\pi^{+}\pi^{-}$ Decay Channel}

For the $\pi^{+}\pi^{-}$ invariant-mass spectrum, the baseline fit model (excluding glueball candidates) consists of the nonresonant DPE continuum and resonance contributions observed in previous measurements, namely $f_{0}(500)$, $f_{0}(980)$, and $f_{2}(1270)$ from DPE, along with $\rho^{0}(770)$ from VMP. The corresponding fit, shown in Appendix~\ref{app:rest_fit_plots}, Fig.~\ref{fig:pipi_baseline}, provides a poor description of the data, yielding
\begin{equation}
    \chi^{2}/\mathrm{ndf}=420.757/61=6.898,
    \label{eq:fit_pipi_baseline_chi2_ndf}
\end{equation}
indicating that the baseline fit model alone is insufficient to reproduce the observed structure in the higher-mass region. The inclusion of the $f_{0}(1500)$ resonance leads to a substantial improvement in the fit quality, as shown in Appendix~\ref{app:rest_fit_plots}, Fig.~\ref{fig:pipi_baseline_1500}, reducing the fit statistic to
\begin{equation}
    \chi^{2}/\mathrm{ndf}=155.542/59=2.636.
    \label{eq:fit_pipi_baseline_w1500_chi2_ndf}
\end{equation}
The inclusion of the additional resonance amplitude improves the fit by $\Delta\chi^{2}=265.2$ for two additional free parameters ($\Delta\mathrm{ndf}=2$), corresponding to a local statistical significance of approximately $16.0\sigma$ (under Wilks' theorem~\cite{Wilks_theorem}).

\begin{figure}[H]
    \includegraphics[width=0.45\textwidth]{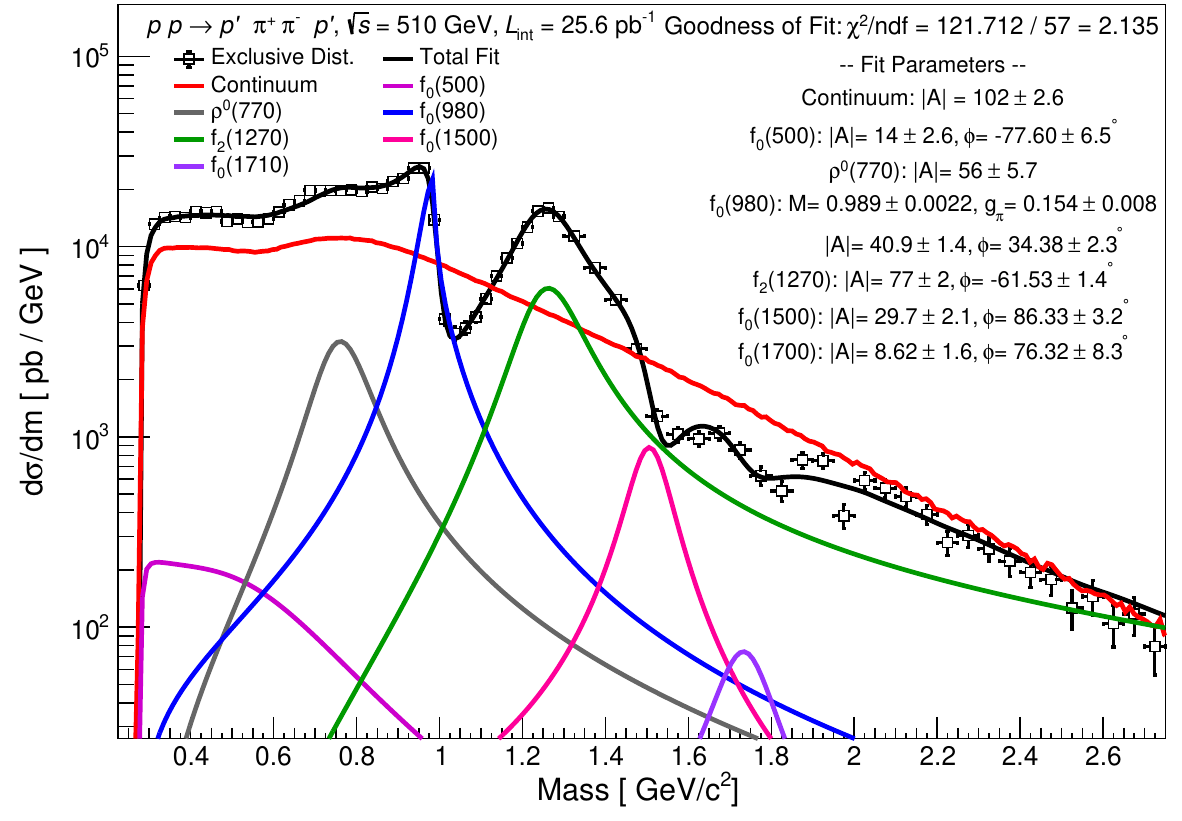}
    \caption{Background-subtracted $\pi^+\pi^-$ differential cross section fitted with the baseline model including the $f_0(1500)$ and $f_0(1710)$ resonances. The fit includes the nonresonant continuum and resonant contributions. The extracted fit parameters and their statistical uncertainties are listed in the figure.}
    \label{fig:pipi_baseline_1500_1710}
\end{figure}

A further improvement is obtained by including the $f_{0}(1710)$ resonance, resulting in the fit shown in Fig.~\ref{fig:pipi_baseline_1500_1710}, with
\begin{equation}
    \chi^{2}/\mathrm{ndf}=121.712/57=2.135.
    \label{eq:fit_pipi_baseline_w1500_1710_chi2_ndf}
\end{equation}
The inclusion of the $f_{0}(1710)$ amplitude improves the fit by $\Delta\chi^{2}=33.830$ for two additional free parameters ($\Delta\mathrm{ndf}=2$), corresponding to a local statistical significance of approximately $5.35\sigma$. 

\begin{figure}[H]
 \includegraphics[width=0.45\textwidth]{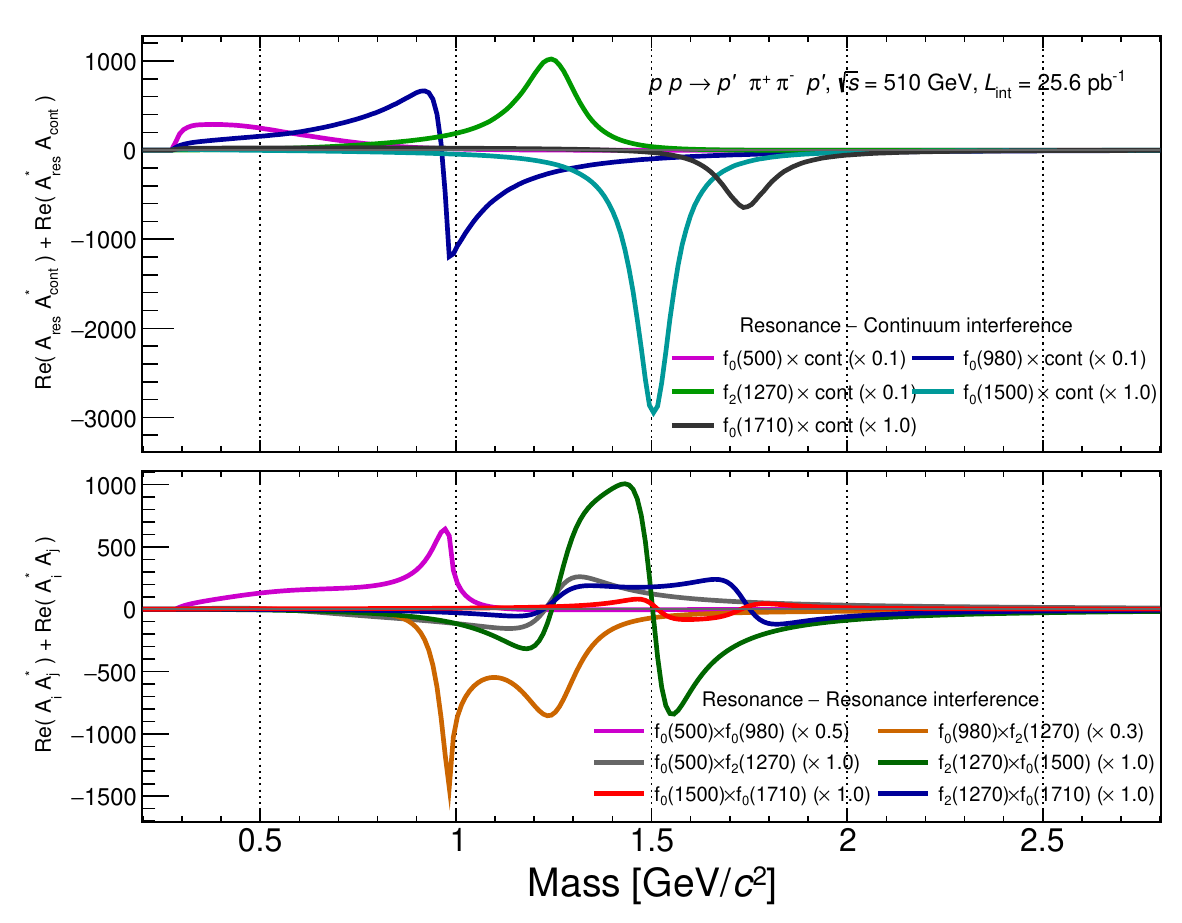}
 \caption{Interference contributions to the $\pi^+\pi^-$ differential cross section for figure~\ref{fig:pipi_baseline_1500_1710}. The top panel shows the interference between the continuum and resonant amplitudes, while the bottom panel shows the interference among different resonant contributions. For clarity, some curves have been scaled for visualization only; the scaling does not affect the fitted results.}
 \label{fig:pipi_baseline_1500_1710_interfernce}
\end{figure}

Although the resulting $\chi^{2}/\mathrm{ndf}$ remains above unity, indicating that the model does not fully describe the observed spectrum within the assigned uncertainties, the remaining discrepancies are predominantly associated with variations in the higher invariant-mass region, particularly above $1.8~\mathrm{GeV}$. While additional resonant contributions could potentially improve the description of this region, the limited statistical precision at higher invariant masses does not provide sufficient sensitivity to reliably constrain their amplitudes. Introducing further resonances would therefore increase the model complexity without adequate statistical justification and could lead to overfitting statistical fluctuations. Consequently, the fit model including the $f_{0}(1710)$ resonance is retained as the final fit configuration for the $\pi^{+}\pi^{-}$ invariant-mass spectrum.

The observed improvement with glueball candidates is not solely due to the addition of extra resonant amplitudes. As illustrated in Fig.~\ref{fig:pipi_baseline_1500_1710_interfernce}, the description of the high-mass region is governed by strong coherent interference between the resonances and the non-resonant continuum, and in between the resonances themselves. Figure~\ref{fig:pipi_baseline_1500_1710_interfernce} also highlights the importance of performing a coherent amplitude analysis rather than treating the resonances as independent incoherent contributions.

The fitted Flatt\'e parameters for the $f_{0}(980)$ resonance are
\begin{align}
\mathrm{M} &= 0.989 \pm 0.002~\mathrm{GeV},\\
g_{\pi} &= 0.154 \pm 0.008~\mathrm{GeV},\,\mathrm{or}\\
\Gamma &= 0.092 \pm 0.005~\mathrm{GeV} \,
\end{align}
and are consistent with those reported by the Particle Data Group (PDG)~\cite{PDG2026}.

The possible contribution of the broad $f_{0}(1370)$ state is particularly noteworthy. The WA102 experiment reported evidence for the $f_{0}(1370)$ in the $\pi^{+}\pi^{-}$ decay channel, with a mass and width of approximately $1308,\mathrm{MeV}$ and $222,\mathrm{MeV}$, respectively~\cite{WA102_pi+pi-}. Owing to its large width and substantial overlap with the prominent $f_{2}(1270)$ resonance, its contribution cannot be readily distinguished from the one-dimensional invariant-mass spectrum alone. Its identification therefore relies on a partial-wave analysis, which separates the scalar and tensor contributions through their distinct angular dependencies. An interesting comparison can be made with radiative $J/\psi$ decays, which provide another gluon-rich production environment. In the partial-wave analysis of the $\pi^{+}\pi^{-}$ spectrum from radiative $J/\psi$ decay, the description of the mass region above $1.0,\mathrm{GeV}$ does not require an $f_{0}(1370)$ contribution in the $S$ wave~\cite{radiative_jpsi_pipi}. The differing observations in these two gluon-rich processes illustrate the challenges associated with establishing the $f_{0}(1370)$. While a possible contribution cannot be excluded in the present data, its presence cannot be established conclusively in this paper. 

\subsection{$K^{+}K^{-}$ Decay Channel}

For the $K^{+}K^{-}$ invariant-mass spectrum, the baseline fit model (excluding glueball candidates) consists of the nonresonant DPE continuum and resonance contributions observed in previous measurements, namely $f_{0}(980)$, $f_{2}(1270)$, and $f_{2}^{\prime}(1525)$ from DPE. The $f_{0}(980)$ Flatt\'e parameters are fixed to the values obtained from the $\pi^{+}\pi^{-}$ analysis, while the masses and widths of the remaining resonances are fixed to their PDG values. The baseline fit, shown in appendix~\ref{app:rest_fit_plots} Fig.~\ref{fig:kk_baseline}, describes the measured spectrum poorly, yielding
\begin{equation}
\chi^{2}/\mathrm{ndf}=92.856/17=5.462
\end{equation}
which clearly indicates that the baseline fit model alone is insufficient. The inclusion of the $f_{0}(1710)$ resonance substantially improves the description of the data, particularly the broad enhancement above the $f_{2}^{\prime}(1525)$ region.

\begin{figure}[H]
 \includegraphics[width=0.45\textwidth]{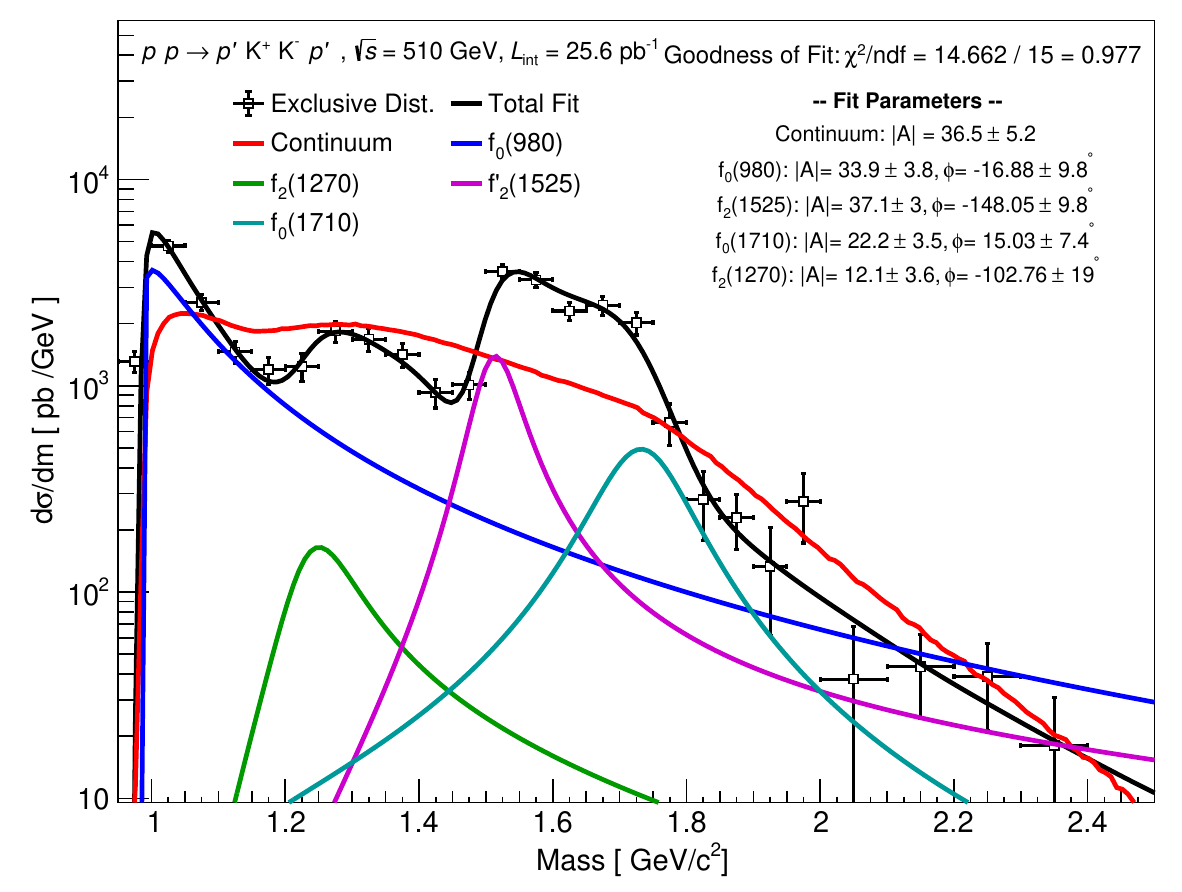}
 \caption{Background-subtracted $K^+K^-$ differential cross section together with the best-fit model. The fit includes the continuum contribution and the resonant components. The extracted fit parameters and their statistical uncertainties are listed in the figure.}
 \label{fig:kk_baseline_1710}
\end{figure}

\begin{figure}[H]
 \includegraphics[width=0.45\textwidth]{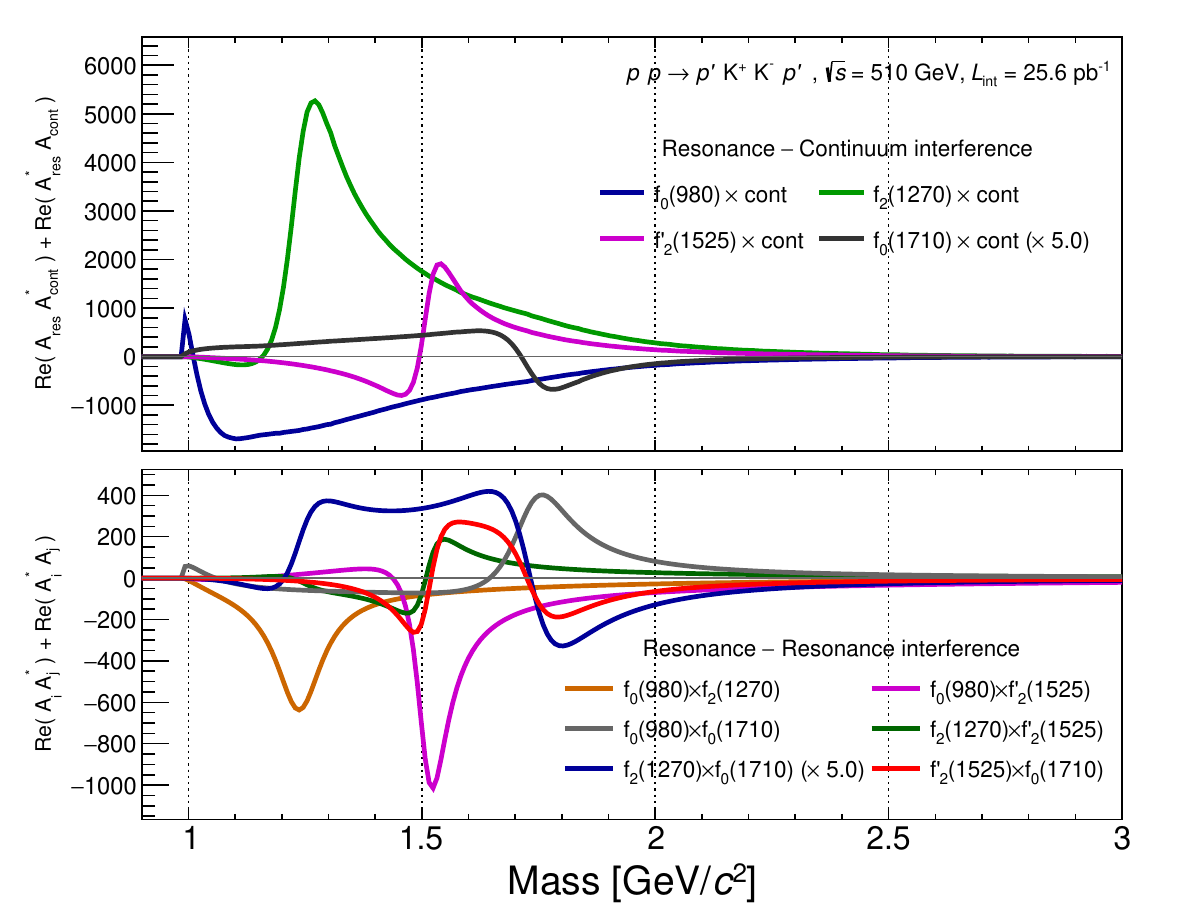}
 \caption{Interference contributions to the $K^+K^-$ differential cross section for figure~\ref{fig:kk_baseline_1710}. The top panel shows the interference between the continuum and resonant amplitudes, while the bottom panel shows the interference among different resonant contributions. For clarity, some curves have been scaled for visualization only; the scaling does not affect the fitted results.}
 \label{fig:kk_baseline_1710_interference}
\end{figure}

The resulting fit, shown in Fig.~\ref{fig:kk_baseline_1710}, gives
\begin{equation}
\chi^{2}/\mathrm{ndf}=14.662/15=0.977
\end{equation}
and improves the fit by $\Delta\chi^{2}=78.194$ for two additional free parameters ($\Delta\mathrm{ndf}=2$), corresponding to a local statistical significance of approximately $8.57\sigma$. Figure~\ref{fig:kk_baseline_1710_interference} demonstrates the importance of treating the amplitudes coherently to account for interference effects as seen in $\pi^+\pi^-$ decay channel. The inclusion of the $f_{0}(1500)$ resonance was also investigated. However, the fit failed to produce a reliable covariance matrix, with \textsc{Minuit} HESSE reporting an unstable error estimate. No quantitative result for the $f_{0}(1500)$ contribution is therefore presented in the $K^{+}K^{-}$ channel.

Although the structure around \(1.5~\mathrm{GeV}/c^{2}\) cannot be conclusively attributed solely to the \(f_{2}^{\prime}(1525)\), with \(f_{0}(1500)\) contribution completely excluded. Previous results from the WA102~\cite{WA102_KKbar} and BES~\cite{radiative_jpsi_kk} Collaborations provide useful guidance. In the \(K^+K^-\) final state, these studies found the \(D\)-wave contribution in this mass region to be substantially more prominent than the corresponding \(S\)-wave contribution, supporting a dominant \(f_{2}^{\prime}(1525)\) component. This interpretation is also consistent with the markedly different \(K\bar{K}\) branching fractions of the two resonances: the \(f_{2}^{\prime}(1525)\) decays to \(K\bar{K}\) with a branching fraction of approximately \(88.8\%\), whereas the corresponding branching fraction for the \(f_{0}(1500)\) is only about \(8.5\%\)~\cite{PDG2026}. Thus, while a small \(f_{0}(1500)\) contribution cannot be ruled out with the present statistics, the observed structure near \(1.5~\mathrm{GeV}/c^{2}\) is naturally expected to be dominated by the \(f_{2}^{\prime}(1525)\).

Similarly, a VMP contribution of $\phi(1020)$ cannot be excluded in the region around $1.02~\mathrm{GeV}$,
which overlaps with the prominent $f_0(980)$ structure. The limited statistics and invariant-mass resolution of the present data do not allow these contributions to be clearly separated. In the review by Kirk~\cite{Kirk_OmegaSpectrometer_Review}, the WA102 measurements give production cross sections of approximately $5711$ and $60~\mathrm{nb}$ for the $f_0(980)$ and $\phi(1020)$, respectively. This large difference suggests that the $f_0(980)$ contribution dominates in central production. However, because the WA102 measurements were performed at a much lower center-of-mass energy and under different experimental conditions, this cross-section ratio cannot be directly applied to the present data. Therefore, although the $f_0(980)$ is expected to provide the dominant contribution in this mass region, a possible contribution from photoproduced $\phi(1020)$ cannot be quantitatively constrained with the present data.

\subsection{Fit Uncertainties}

A detailed evaluation of the systematic uncertainties is provided in Ref.~\cite{STARCEP_510GeV}. The background-subtraction uncertainty is the only bin-dependent contribution and is below $1\%$; it is therefore neglected in the present analysis. The remaining systematic uncertainties are bin-independent and mainly affect the overall normalization of the spectra. They are therefore correlated across the invariant-mass bins. Treating these uncertainties as independent bin-by-bin contributions could distort the $\chi^2$ and the fitted parameter uncertainties.

\begin{table}[H]
\centering
\caption{Fitted amplitudes and phases for the $\pi^{+}\pi^{-}$ channel with their statistical and systematic uncertainties from varying Orear continuum parameters}
\label{tab:fit_parameters_pion}
\begin{tabular}{lccc}
\toprule
\makecell{Parameter\\ A [$\sqrt{pb/GeV}$], $\phi$ [deg]} & \makecell{Nominal\\Value} & \makecell{Statistical\\Uncertainity} & \makecell{Model\\Uncertainty} \\
\midrule
$\chi^2 /ndf$             & 2.13   & --   & 0.17 \\ \\
$|A_{\mathrm{cont}}|$     & 102.37 & 2.55 & 1.34 \\
$|A_{f_{0}(500)}|$        & 14.04  & 2.62 & 2.27 \\
$|A_{\rho}|$              & 56.02  & 5.70 & 3.25 \\
$|A_{f_{0}(980)}|$        & 40.93  & 1.37 & 1.16 \\
$|A_{f_{2}(1270)}|$       & 76.96  & 2.00 & 0.68 \\
$|A_{f_{0}(1500)}|$       & 29.66  & 2.13 & 0.44 \\
$|A_{f_{0}(1710)}|$       & 8.62   & 1.56 & 0.20 \\ \\
$\phi_{f_{0}(500)}$       & -1.35  & 0.11 & 0.03 \\
$\phi_{f_{0}(980)}$       & 0.60   & 0.04 & 0.02 \\
$\phi_{f_{2}(1270)}$      & -1.07  & 0.02 & 0.05 \\
$\phi_{f_{0}(1500)}$      & 1.51   & 0.06 & 0.01 \\
$\phi_{f_{0}(1710)}$      & 1.33   & 0.14 & 0.09 \\ \\
$M(f_{0}(980))$ [MeV]     & 988.66 & 2.16 & 0.37 \\
$g_{\pi}(f_0(980))$       & 0.154  & 0.008 & 0.006 \\
\bottomrule
\end{tabular}
\end{table}

  Another source of uncertainty in this analysis arises from the modeling of the continuum contribution. The continuum distribution depends on the off-shell meson form factor, which is parameterized by two parameters, $a_{\mathrm{OREAR}} : 0.45 \, \mathrm{GeV}$ and $b_{\mathrm{OREAR}} : 1.37 \, \mathrm{GeV^{-1}}$~\cite{Graniitti}. To estimate the associated systematic uncertainty, each parameter was independently varied by $\pm10\%$, and the fit was repeated for each variation. The resulting model uncertainties on the free fit parameters were evaluated from the deviations with respect to the nominal fit according to Eq.~(\ref{eq:syst_unc}) and are summarized in Tables~\ref{tab:fit_parameters_pion} and~\ref{tab:fit_parameters_kaon}. For both the $\pi^{+}\pi^{-}$ and $K^{+}K^{-}$ decay channels, the estimated model uncertainties are generally smaller than the corresponding statistical uncertainties.

\begin{table}[htbp]
\centering
\caption{Fitted amplitudes and phases for the $K^{+}K^{-}$ channel with their statistical and systematic uncertainties from varying Orear continuum parameters}
\label{tab:fit_parameters_kaon}
\begin{tabular}{lccc}
\toprule
\makecell{Parameter\\ A [$\sqrt{pb/GeV}$], $\phi$ [deg]} & \makecell{Nominal\\Value} & \makecell{Statistical\\Uncertainity} & \makecell{Model\\Uncertainty} \\
\midrule
$\chi^2 /ndf$             & 0.98  & --   & 0.03 \\ \\
$|A_{\mathrm{cont}}|$     & 36.53 & 5.24 & 0.82 \\
$|A_{f_{0}(980)}|$        & 33.89 & 3.80 & 1.34 \\
$|A_{f_{2}(1270)}|$       & 12.07 & 3.55 & 0.15 \\
$|A_{f_{2}(1525)}|$       & 37.11 & 2.99 & 0.70 \\
$|A_{f_{0}(1710)}|$       & 22.19 & 3.51 & 0.41 \\ \\
$\phi_{f_{0}(980)}$       & -0.29 & 0.17 & 0.05 \\
$\phi_{f_{2}(1270)}$      & -1.79 & 0.33 & 0.08 \\
$\phi_{f_{2}(1525)}$      & -2.58 & 0.17 & 0.02 \\
$\phi_{f_{0}(1710)}$      &  0.26 & 0.13 & 0.04 \\
\bottomrule
\end{tabular}
\end{table}


\section{Fiducial Cross-Section Measurement}

The fitted mass distributions provide differential contributions as a function of the invariant mass \(m\). The corresponding fiducial cross-section for a given interference component \(c\) is obtained by integrating its distribution over the fiducial mass range,

\begin{equation}
\sigma_c
=
\int_{m_{\min}}^{m_{\max}}
F_c(m)\,dm,
\label{eq:component_cross_section}
\end{equation}

where \(F_c(m)\) denotes the fitted distribution associated with component \(c\). Depending on the component under consideration, \(F_c(m)\) may represent a self contribution, an interference contribution, or the total fitted distribution. The limits \(m_{\min}\) and \(m_{\max}\) are 0 and 3.0 GeV.

Since the fitted distributions are evaluated at discrete mass points, the integrals are evaluated numerically using the trapezoidal rule. For a distribution sampled at mass points \(m_k\), the fiducial cross-section is therefore calculated as

\begin{equation}
\sigma_c
\simeq
\sum_{k=1}^{n-1}
\frac{
F_c(m_k)+F_c(m_{k+1})
}{2}
\left(
m_{k+1}-m_k
\right)
\label{eq:graph_integral}
\end{equation}

For the coherent DPE contribution, particular care is required when integrating the interference terms, which arise from products of different amplitudes. These terms are integrated with their physical signs, such that positive interference increases the integrated cross-section, whereas negative interference reduces it. Consequently, the integrated cross-section cannot, in general, be obtained by simply summing the individual self contributions, as interference terms may partially cancel over the fiducial mass range. The quoted fiducial cross-sections therefore retain the full interference structure of the fitted amplitudes, including their relative phases, ensuring that they represent the physical coherent sum rather than an incoherent sum of individual contributions.

\subsection{$\pi^{+}\pi^{-}$ Decay Channel}

The total cross section obtained using the final fit function, as presented in Table~\ref{tab:fit_xsecs_pion}, is consistent with the total cross section reported in the published measurement~\cite{STARCEP_510GeV} within statistical uncertainties.


\begin{table}[H]
\centering
\caption{Fit graph integrated fiducial cross sections.}
\label{tab:fit_xsecs_pion}
\begin{tabular}{lrrr}
\toprule
\makecell{ $\sigma^{\mathrm{fid}}$ (pb) } &
\makecell{Nominal\\Value } &
\makecell{Statistical\\Uncertainty } &
\makecell{Model\\Uncertainty } \\
\midrule
Fit total
    & 16943.98 & 78.90 & 5.01 \\
Continuum
    & 10452.01 & 521.26 & 283.84 \\
$f_{0}(500)$
    & 93.00 & 34.75 & 28.53 \\
$\rho^{0}(770)$
    & 686.21 & 139.73 & 79.24 \\
$f_{0}(980)$
    & 1853.28 & 92.04 & 88.68 \\
$f_{2}(1270)$
    & 1681.39 & 87.35 & 29.46 \\
$f_{0}(1500)$
    & 151.80 & 21.78 & 4.50 \\
$f_{0}(1710)$
    & 17.09 & 6.19 & 0.79 \\
\midrule
$f_{0}(500)\times\mathrm{cont}$
    & 1080.51 & 237.96 & 114.44 \\
$f_{0}(980)\times\mathrm{cont}$
    & -509.77 & 135.41 & 130.67 \\
$f_{2}(1270)\times\mathrm{cont}$
    & 2568.75 & 61.38 & 51.76 \\
$f_{0}(1500)\times\mathrm{cont}$
    & -497.09 & 52.05 & 12.10 \\
$f_{0}(1710)\times\mathrm{cont}$
    & -108.33 & 41.99 & 16.63 \\
$f_{0}(500)\times f_{0}(980)$
    & 283.13 & 48.14 & 59.24 \\
$f_{0}(500)\times f_{2}(1270)$
    & 43.29 & 22.84 & 10.46 \\
$f_{0}(980)\times f_{2}(1270)$
    & -1155.24 & 68.57 & 30.12 \\
$f_{2}(1270)\times f_{0}(1500)$
    & -81.51 & 34.32 & 29.00 \\
$f_{2}(1270)\times f_{0}(1710)$
    & 36.67 & 18.50 & 19.33 \\
$f_{0}(1500)\times f_{0}(1710)$
    & 15.76 & 7.32 & 3.70 \\
\bottomrule
\end{tabular}
\end{table}

 The dominant contribution to the total cross section arises from the non-resonant continuum, followed by the contributions from the \(f_0(980)\) and \(f_2(1270)\) resonances. The interference terms exhibit both constructive and destructive contributions, demonstrating their non-negligible impact on the overall cross section. In particular, the \(f_0(980)\times f_2(1270)\) interference term provides the largest negative contribution, whereas the interference between the \(f_2(1270)\) resonance and the non-resonant continuum gives the dominant positive contribution. A detailed decomposition of the measured cross section into the individual resonant, non-resonant, and interference components provides additional information on the underlying production dynamics and can serve as an important constraint for future theoretical models of the process.


\begin{table}[H]
\centering
\caption{Resonance self-term ratios with respect to the fit total.}
\label{tab:ratios_total_pipi}
\begin{tabular}{lrrr}
\toprule
\makecell{ $\sigma^{\mathrm{fid}}_{1}/\sigma^{\mathrm{fid}}_{2}$ ($\times 0.01$) } &
\makecell{Nominal\\Value } &
\makecell{Statistical\\Uncertainty } &
\makecell{Model\\Uncertainty } \\
\midrule
$f_{0}(500)/\mathrm{total}$       & 0.55 & 0.21 & 0.17 \\
$\rho^{0}(770)/\mathrm{total}$    & 4.05 & 0.82 & 0.47 \\
$f_{0}(980)/\mathrm{total}$       & 10.94 & 0.55 & 0.53 \\
$f_{2}(1270)/\mathrm{total}$      & 9.92 & 0.52 & 0.18 \\
$f_{0}(1500)/\mathrm{total}$      & 0.90 & 0.13 & 0.03 \\
$f_{0}(1710)/\mathrm{total}$      & 0.10 & 0.04 & 0.00 \\
\bottomrule
\end{tabular}
\end{table}

The measured fiducial cross sections are affected by the detector acceptance and the kinematic requirements imposed on the final-state particles. Consequently, their absolute values cannot be directly compared with measurements performed in other decay channels or by other experiments with different fiducial acceptances. A useful complementary quantity is therefore the ratio of the individual cross-section contributions to the total cross section obtained from the fit. These ratios, presented in Table~\ref{tab:ratios_total_pipi}, provide a normalized characterization of the relative contributions of the different components to the measured cross section and facilitate comparisons with other measurements and theoretical predictions.


\begin{table}[H]
\centering
\caption{Ratio of the $f_{0}(1500)$ and $f_{0}(1710)$ self terms.}
\label{tab:ratio_f01500_f01710}
\begin{tabular}{lrrr}
\toprule
\makecell{ $\sigma^{\mathrm{fid}}_{1}/\sigma^{\mathrm{fid}}_{2}$ } &
\makecell{Nominal\\Value } &
\makecell{Statistical\\Uncertainty } &
\makecell{Model\\Uncertainty } \\
\midrule
$f_{0}(1500)/f_{0}(1710)$ & 8.88 & 3.46 & 0.16 \\
\bottomrule
\end{tabular}
\end{table}

Another quantity of interest is the cross-section ratio between the glueball candidates \(f_0(1500)\) and \(f_0(1710)\), given in Table~\ref{tab:ratio_f01500_f01710}. The measured ratio of \(8.88\) indicates a significantly larger \(f_0(1500)\) contribution in this final state within the measured fiducial region.

\subsection{$K^{+}K^{-}$ Decay Channel}


\begin{table}[H]
\centering
\caption{Fit graph integrated fiducial cross sections.}
\label{tab:fit_xsecs_kaon}
\begin{tabular}{lrrr}
\toprule
\makecell{ $\sigma^{\mathrm{fid}}$ (pb) } &
\makecell{Nominal\\Value } &
\makecell{Statistical\\Uncertainty } &
\makecell{Model\\Uncertainty } \\
\midrule
Fit total
    & 1677.92 & 48.42 & 2.60 \\
Continuum
    & 1334.28 & 382.48 & 59.93 \\
$f_{0}(980)$
    & 630.28 & 141.36 & 51.12 \\
$f_{2}(1270)$
    & 41.46 & 24.46 & 1.04 \\
$f_{2}(1525)$
    & 176.38 & 28.42 & 6.63 \\
$f_{0}(1710)$
    & 110.90 & 35.10 & 4.09 \\
\midrule
$f_{0}(980)\times\mathrm{cont}$
    & -924.93 & 425.27 & 113.23 \\
$f_{2}(1270)\times\mathrm{cont}$
    & 288.03 & 63.35 & 7.17 \\
$f_{2}(1525)\times\mathrm{cont}$
    & 150.65 & 39.16 & 15.90 \\
$f_{0}(1710)\times\mathrm{cont}$
    & 124.53 & 41.86 & 14.31 \\
$f_{0}(980)\times f_{2}(1270)$
    & -167.11 & 49.99 & 3.48 \\
$f_{0}(980)\times f_{2}(1525)$
    & -157.44 & 20.74 & 5.36 \\
$f_{0}(980)\times f_{0}(1710)$
    & 69.90 & 43.65 & 6.53 \\
$f_{2}(1270)\times f_{2}(1525)$
    & -5.36 & 19.18 & 4.09 \\
$f_{2}(1270)\times f_{0}(1710)$
    & 32.08 & 17.42 & 2.40 \\
$f_{2}(1525)\times f_{0}(1710)$
    & -29.08 & 23.47 & 2.64 \\
\bottomrule
\end{tabular}
\end{table}

Similarly, the total cross section obtained from the final fit, presented in Table~\ref{tab:fit_xsecs_kaon}, is consistent with the published measurement~\cite{STARCEP_510GeV} within statistical uncertainties. The dominant contribution arises from the non-resonant continuum, followed by the \(f_0(980)\) and \(f_2(1525)\) resonances. The largest negative interference contribution originates from the \(f_0(980)\times\) non-resonant continuum term, while the \(f_2(1270)\times\) non-resonant continuum term provides the largest positive contribution. The ratios of the individual resonance contributions to the total fit are summarized in Table~\ref{tab:ratios_total_kaon}.


\begin{table}[H]
\centering
\caption{Resonance self-term ratios with respect to the fit total.}
\label{tab:ratios_total_kaon}
\begin{tabular}{lrrr}
\toprule
\makecell{ $\sigma^{\mathrm{fid}}_{1}/\sigma^{\mathrm{fid}}_{2}$ ($\times 0.01$) } &
\makecell{Nominal\\Value } &
\makecell{Statistical\\Uncertainty } &
\makecell{Model\\Uncertainty } \\
\midrule
$f_{0}(980)/\mathrm{total}$
    & 37.56 & 8.49 & 3.10 \\
$f_{2}(1270)/\mathrm{total}$
    & 2.47 & 1.46 & 0.06 \\
$f_{2}(1525)/\mathrm{total}$
    & 10.51 & 1.72 & 0.38 \\
$f_{0}(1710)/\mathrm{total}$
    & 6.61 & 2.10 & 0.25 \\
\bottomrule
\end{tabular}
\end{table}

Assuming isospin symmetry, \(\mathcal{B}(f_{0}\to K^{+}K^{-})=\frac{1}{2}\mathcal{B}(f_{0}\to K\bar{K})\) and \(\mathcal{B}(f_{0}\to\pi^{+}\pi^{-})=\frac{2}{3}\mathcal{B}(f_{0}\to\pi\pi)\). Branching fraction ratios measured by WA102 CEP ~\cite{WA102_coupled_channel} correspond to
\begin{align}
\frac{\mathcal{B}(f_{0}(1500)\to K^{+}K^{-})}
     {\mathcal{B}(f_{0}(1500)\to \pi^{+}\pi^{-})}
&\simeq 0.25, \\
\frac{\mathcal{B}(f_{0}(1710)\to K^{+}K^{-})}
     {\mathcal{B}(f_{0}(1710)\to \pi^{+}\pi^{-})}
&\simeq 3.75.
\end{align}

 This behavior is also qualitatively consistent with the measured fiducial cross sections. In particular, the ratio of the \(f_{0}(1710)\) contribution relative to the total fiducial cross section in the \(K^{+}K^{-}\) channel to that in the \(\pi^{+}\pi^{-}\) channel is approximately \(6.0\). Although an explicit \(f_{0}(1500)\) contribution is not reported in the \(K^{+}K^{-}\) channel, its possible magnitude can be estimated indirectly using the charged-mode branching-fraction ratio of 0.25. Taking the measured \(f_{0}(1500)\) contribution in the \(\pi^{+}\pi^{-}\) channel to be approximately \(0.9\) of the corresponding total fiducial cross section would therefore suggest an expected relative contribution of approximately $ 0.25\times0.9\simeq0.225 $ in the \(K^{+}K^{-}\) channel. On comparing this value with other resonances from table~\ref{tab:ratios_total_kaon}, shows the relatively small contribution in this channel which makes it challenging to find in vicinity of $f_2(1525)$.

\section{Conclusion}

A coherent amplitude analysis of the \(\pi^{+}\pi^{-}\) and \(K^{+}K^{-}\) invariant-mass spectra has been performed using the published STAR central exclusive production (CEP) dataset from \(pp\) collisions at \(\sqrt{s}=510~\mathrm{GeV}\). In the \(\pi^{+}\pi^{-}\) channel, the baseline fit model, comprising the nonresonant double-Pomeron exchange (DPE) continuum and the \(f_{0}(500)\), \(\rho^{0}(770)\), \(f_{0}(980)\), and \(f_{2}(1270)\) resonances, does not adequately describe the measured spectrum above the \(f_{2}(1270)\) region. The successive inclusion of the \(f_{0}(1500)\) and \(f_{0}(1710)\) amplitudes leads to substantial improvements in the fit quality, demonstrating the importance of both contributions in describing the observed invariant-mass spectrum. Similarly, in the \(K^{+}K^{-}\) channel, the inclusion of the \(f_{0}(1710)\) amplitude significantly improves the baseline fit model, which consists of the nonresonant DPE continuum and the \(f_{0}(980)\), \(f_{2}(1270)\), and \(f_{2}^{\prime}(1525)\) resonances. These results provide quantitative evidence for contributions associated with both leading scalar-glueball candidates, \(f_{0}(1500)\) and \(f_{0}(1710)\), in CEP at the highest collision energy of RHIC \(pp\).

The present analysis does not, by itself, determine the underlying quark or gluonic composition of the scalar states. Nevertheless, the observed channel dependence provides information that can be considered in the context of previous measurements and phenomenological interpretations. In particular, the comparatively prominent $f_{0}(1710)$ contribution in the $K^{+}K^{-}$ channel, approximately six times larger than in the $\pi^{+}\pi^{-}$ channel in the present analysis, is qualitatively consistent with its known preference for decays to $K\bar K$ over $\pi\pi$, as observed in previous studies, including low-energy WA102 CEP~\cite{WA102_coupled_channel}, radiative $J/\psi$ decays~\cite{JinShen2021}, and two-photon measurements~\cite{Belle_KSKS_2013}. Interestingly, this behavior can be accommodated within different glueball--quarkonium mixing scenarios for different reasons. In the CKZ model, in which the $f_{0}(1710)$ is predominantly $s\bar{s}$, its enhanced contribution to the $K\bar K$ channel naturally follows from its large strange-quark content. In contrast, in the CCL model the $f_{0}(1710)$ is predominantly gluonic; within this interpretation, chiral suppression of scalar-glueball decays suppresses the coupling to light quarks relative to strange quarks, leading to an enhanced $K\bar K$ decay rate relative to $\pi\pi$. Thus, the stronger $f_{0}(1710)$ contribution observed in the $K^{+}K^{-}$ channel is qualitatively compatible with both mixing scenarios and, by itself, does not discriminate between an $s\bar{s}$-dominated and a glueball-dominated interpretation. 

The $f_{0}(1500)$ is clearly observed in the $\pi^{+}\pi^{-}$ spectrum, whereas no significant contribution is required in the $K^{+}K^{-}$ spectrum. When the relative decay rates measured by the WA102 CEP experiment are imposed, the expected $f_{0}(1500)$ contribution to the $K^{+}K^{-}$ channel is smaller than those of the other resonances included in the fit. An important measure of the relative contributions of the two scalar glueball candidates in the $\pi^{+}\pi^{-}$ channel is the ratio of their fractional fitted cross sections,

\[
\frac{
\left[\sigma_{f_{0}(1500)}/\sigma_{\mathrm{total\,fit}}\right]_{\pi^{+}\pi^{-}}
}{
\left[\sigma_{f_{0}(1710)}/\sigma_{\mathrm{total\,fit}}\right]_{\pi^{+}\pi^{-}}
}
= 8.8 \pm 3.46\,(\mathrm{stat.}) \pm 0.16\,(\mathrm{mod.})
\]

Thus, within the present analysis, the relative contribution of the $f_{0}(1500)$ to the $\pi^{+}\pi^{-}$ final state is substantially larger than that of the $f_{0}(1710)$. This statement is consistent with WA102 CEP and radiative $J/\psi$ measurements~\cite{WA102_coupled_channel, radiative_jpsi_pipi} and both the CKZ and CCL modeling.

The present results provide further evidence for the decay-channel dependence of the $f_{0}(1500)$ and $f_{0}(1710)$ resonance contributions in CEP. At $\sqrt{s}=510~\mathrm{GeV}$, the highest collision energy at which CEP has been studied with tagged forward protons, the observed pattern is qualitatively consistent with that reported by the WA102 fixed-target experiment at much lower energies: the $f_{0}(1500)$ contributes more strongly to the $\pi^{+}\pi^{-}$ channel, while the $f_{0}(1710)$ contributes more strongly to the $K^{+}K^{-}$ channel. The observation of this pattern in both RHIC collider and WA102 fixed-target measurements, despite substantial differences in collision energies and detector acceptances, strengthens the experimental evidence for their distinct decay-channel behavior. These results provide important experimental input for understanding the nature of these scalar-isoscalar states and assessing their possible gluonic content and mixing with quark-antiquark components.


\begin{acknowledgments}

The authors would like to thank the STAR Collaboration for providing the invariant-mass resolution obtained from the \(\pi^{+}\pi^{-}\) and \(K^{+}K^{-}\) embedding samples (see Appendix~\ref{app:mass_resolution}), which were used to study the detector efficiency and acceptance for CEP in \(pp\) collisions at \(\sqrt{s}=500~\mathrm{GeV}\). The work presented here is supported by the U.S. Department of Energy, Office of Science, under Award Number DE-FG02-07ER41521.

\end{acknowledgments}

\appendix


\section{Uncertainty Calculations}
\label{app:model_uncertainity}
The systematic uncertainty for each fitted parameter was evaluated as the weighted root-mean-square deviation of the values obtained from the systematic variations with respect to the nominal fit,

\begin{equation}
\sigma_{\mathrm{syst}}
=
\sqrt{
\frac{\displaystyle\sum_{i} w_i \left(x_i-x_{\mathrm{nom}}\right)^2}
{\displaystyle\sum_{i} w_i}
},
\label{eq:syst_unc}
\end{equation}

where $x_{\mathrm{nom}}$ is the value obtained from the nominal fit, $x_i$ is the value obtained from the $i$-th systematic variation, and the weight is defined as

\begin{equation}
w_i=
\begin{cases}
1/\sigma_{\mathrm{stat},i}^{2}, & \sigma_{\mathrm{stat},i}\neq 0,\\[4pt]
1, & \text{if }\sigma_{\mathrm{stat},i}\text{ is zero or unavailable},
\end{cases}
\label{eq:weight}
\end{equation}

where $\sigma_{\mathrm{stat},i}$ is the statistical uncertainty of the corresponding fit parameter obtained from the $i$-th systematic variation.

The statistical uncertainty in the fiducial cross-section measurement was evaluated by propagating the uncertainties and correlations of the fitted parameters:

$$
\delta\sigma_{c,\mathrm{stat}}^2
=
\sum_i
\left(
\frac{\partial\sigma_c}{\partial p_i}
\right)^2
\sigma_i^2
+
2\sum_{i<j}
\frac{\partial\sigma_c}{\partial p_i}
\frac{\partial\sigma_c}{\partial p_j}
\rho_{ij}\sigma_i\sigma_j ,
\label{eq:expanded_covariance_error}
$$

where \(p_i\) denotes the fit parameters relevant to the decay channel under consideration, \(\sigma_i\) is the uncertainty associated with parameter \(p_i\), and \(\rho_{ij}\) is the correlation coefficient between parameters \(p_i\) and \(p_j\).

The first term represents the contributions from the individual parameter uncertainties, while the second term accounts for the covariance between different fit parameters. The correlation matrix of the fit parameters is obtained directly from the fit and is presented in Appendix~\ref{app:correlation}.

\section{Resonance Line Shapes and Mass Resolution}
\label{app:resonance_models}

This appendix summarizes the resonance line shapes and the
mass-resolution parameterization used in the analysis. The invariant
mass is denoted by $m$, while $M$ represents the nominal resonance
mass. Natural units, $\hbar=c=1$, are used throughout.

\subsection{Relativistic Breit--Wigner parameterization}
\label{app:breit_wigner}

A relativistic Breit--Wigner amplitude with a mass-dependent width is
used for resonances decaying into two pions:
\begin{equation}
\mathcal{A}_{\mathrm{BW}}(m)
=
\frac{M\sqrt{\Gamma_0\,\Gamma(m)}}
     {M^2-m^2-iM\Gamma(m)} ,
\label{eq:bw_amplitude}
\end{equation}
where $\Gamma_0$ is the nominal width of the resonance and
$\Gamma(m)$ is its mass-dependent width. The normalization appearing
in the numerator is a convention adopted in the implementation and
does not affect the pole structure of the amplitude.

For a $\pi^+\pi^-$ final state (similarly for $K^+K^-$ using $m_K=0.493677~\mathrm{GeV}$), the breakup momentum in the resonance
rest frame is
\begin{equation}
q(m)
=
\frac{1}{2}
\sqrt{\max\!\left(0,m^2-4m_\pi^2\right)},
\qquad
q_0=q(M),
\label{eq:pion_breakup_momentum}
\end{equation}
with
\begin{equation}
m_\pi=0.13957~\mathrm{GeV}.
\end{equation}
The mass-dependent width is given by
\begin{equation}
\Gamma(m)
=
\Gamma_0
\left(\frac{q(m)}{q_0}\right)^{2L+1}
\frac{M}{m}
\left[
\frac{B_L\!\left(q(m)\right)}
     {B_L(q_0)}
\right]^2 ,
\label{eq:mass_dependent_width}
\end{equation}
where $L$ is the orbital angular momentum of the decay. The factor
$(q/q_0)^{2L+1}$ describes the threshold behavior, $M/m$ is the
relativistic phase-space correction, and $B_L$ is the
Blatt--Weisskopf barrier factor. By construction,
$\Gamma(M)=\Gamma_0$.

The barrier factors are evaluated using
\begin{equation}
z=(qR)^2,
\qquad
R=1~\mathrm{GeV}^{-1},
\end{equation}
and are defined as
\begin{equation}
B_L(q)=
\begin{cases}
1,
& L=0,\\[6pt]
\dfrac{1}{\sqrt{1+z}},
& L=1,\\[12pt]
\dfrac{1}{\sqrt{z^2+3z+9}},
& L=2.
\end{cases}
\label{eq:blatt_weisskopf}
\end{equation}
These choices correspond to $S$-, $P$-, and $D$-wave decays,
respectively. The amplitude is set to zero when $q_0=0$ or $m\leq 0$.

\subsection{Flatté parameterization}
\label{app:flatte}

For a resonance located near the $K\bar{K}$ threshold, a coupled-channel
Flatté-like amplitude is used:
\begin{equation}
\mathcal{A}_{\mathrm{F}}(m)
=
\frac{M\sqrt{\Gamma_{\pi}(m)}}
     {M^2-m^2
      -iM\left[\Gamma_{\pi}(m)+\Gamma_{K}(m)\right]} .
\label{eq:flatte_amplitude}
\end{equation}
The mass-dependent contributions from the $\pi\pi$ and $K\bar{K}$
channels are
\begin{align}
\Gamma_{\pi}(m)
&=g_{\pi}\,q_{\pi}(m),
\\
\Gamma_{K}(m)
&=g_{\pi}
  \left(\frac{g_K}{g_\pi}\right)q_K(m),
\label{eq:flatte_widths}
\end{align}
where the relative coupling is fixed to
\begin{equation}
\frac{g_K}{g_\pi}=4.21.
\label{eq:flatte_coupling_ratio}
\end{equation}
The breakup momentum for a channel containing two particles of mass
$m_j$ is implemented as
\begin{equation}
q_j(m)
=
\begin{cases}
\dfrac{1}{2}\sqrt{m^2-4m_j^2},
& m>2m_j,\\[8pt]
0,
& m\leq 2m_j.
\end{cases}
\label{eq:flatte_channel_momentum}
\end{equation}
The particle masses used in the calculation are
\begin{equation}
m_\pi=0.13957~\mathrm{GeV},
\qquad
m_K=0.493677~\mathrm{GeV}.
\end{equation}
The total width entering the denominator can therefore be written as
\begin{equation}
\Gamma_{\mathrm{tot}}(m)
=
g_\pi
\left[
q_\pi(m)+4.21\,q_K(m)
\right].
\label{eq:flatte_total_width}
\end{equation}
 $\Gamma_{\mathrm{tot}}(m)$ is also effective $f_0(980)$ width. The coupling ratio $g_{K}/g_{\pi}$ is fixed to 4.21, as determined by the BES Collaboration~\cite{Ablikim2005Resonances} and subsequently adopted in the STAR analysis at $\sqrt{s}=200$ GeV~\cite{STAR_CEP_200GeV}.


\subsection{Invariant-mass resolution}
\label{app:mass_resolution}

The invariant-mass resolution of the STAR detector is parameterized using the empirical mass-dependent function
\begin{equation}
\sigma_{\mathrm{res}}(M)
=
a+\frac{b}{M}+\frac{c}{M^2}+dM+eM^2,
\label{eq:mass_resolution_general}
\end{equation}
where \(M\) denotes the invariant mass of the two-particle system and \(a\), \(b\), \(c\), \(d\), and \(e\) are parameters determined from a fit to the mass-resolution distributions.

\begin{figure}[H]
\centering
\includegraphics[width=0.80\linewidth]{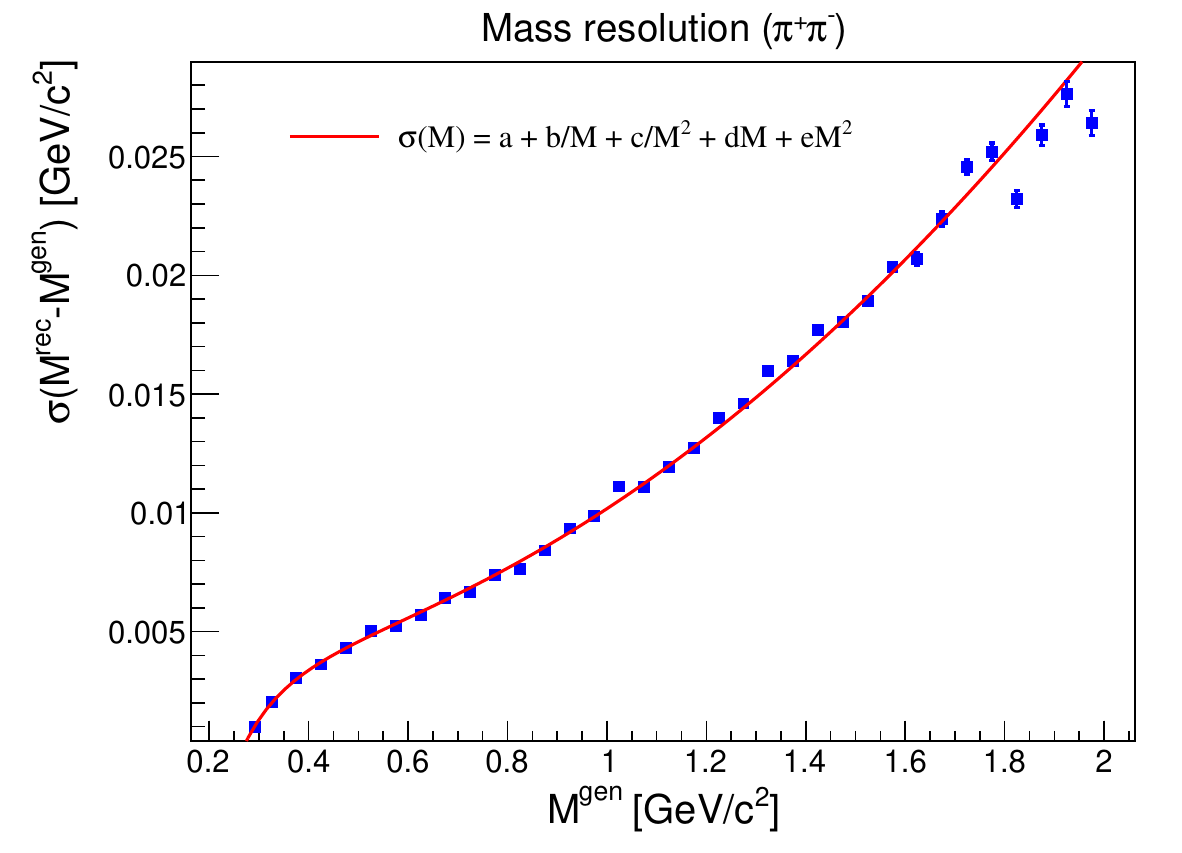}
\caption{Invariant-mass resolution for the \(\pi^{+}\pi^{-}\) final state obtained from the STAR embedding sample at \(\sqrt{s}=510~\mathrm{GeV}\). The mass-dependent parameterization of Eq.~\eqref{eq:mass_resolution_general} is used to describe the resolution.}
\label{fig:pipi_resol}
\end{figure}

\begin{figure}[H]
\centering
\includegraphics[width=0.80\linewidth]{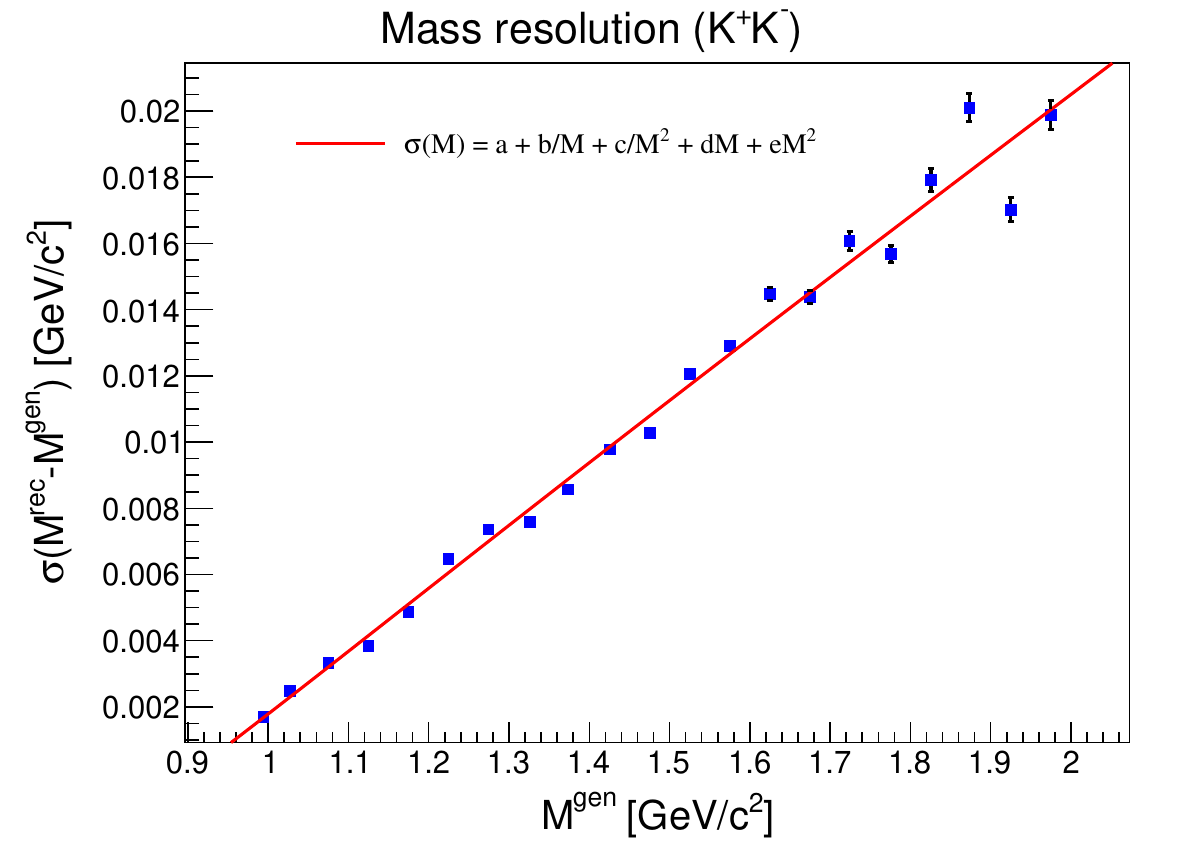}
\caption{Invariant-mass resolution for the \(K^{+}K^{-}\) final state obtained from the STAR embedding sample at \(\sqrt{s}=510~\mathrm{GeV}\). The mass-dependent parameterization of Eq.~\eqref{eq:mass_resolution_general} is used to describe the resolution.}
\label{fig:kk_resol}
\end{figure}

The invariant-mass resolution was determined using STAR embedding samples corresponding to \(pp\) collisions at \(\sqrt{s}=510~\mathrm{GeV}\). These samples, containing embedded \(\pi^{+}\pi^{-}\) and \(K^{+}K^{-}\) pairs, were originally produced for detector efficiency and acceptance studies. The reconstructed track pairs were subjected to the same track selection criteria as those applied to the experimental data~\cite{STARCEP_510GeV}, except for the Roman Pot selection requirements, which were not relevant to the central-track efficiency calculations. For each final state, the invariant-mass resolution was evaluated as a function of the generated two-particle invariant mass. The resulting mass-resolution points, together with their corresponding parameterizations, are presented in Figs.~\ref{fig:pipi_resol} and~\ref{fig:kk_resol} for the \(\pi^{+}\pi^{-}\) and \(K^{+}K^{-}\) final states, respectively. It should be noted that the invariant-mass resolution depends not only on the invariant mass of the particle pair but also on the transverse momentum (\(p_T\)) and pseudorapidity (\(\eta\)) distributions of the daughter particles. Consequently, the reported mass resolution is specific to the kinematic phase space defined by the track selection and particle identification criteria adopted in the published STAR analysis~\cite{STARCEP_510GeV}.
\section{Fit Plots}
\label{app:rest_fit_plots}

\begin{figure}[H]
 \includegraphics[width=0.45\textwidth]{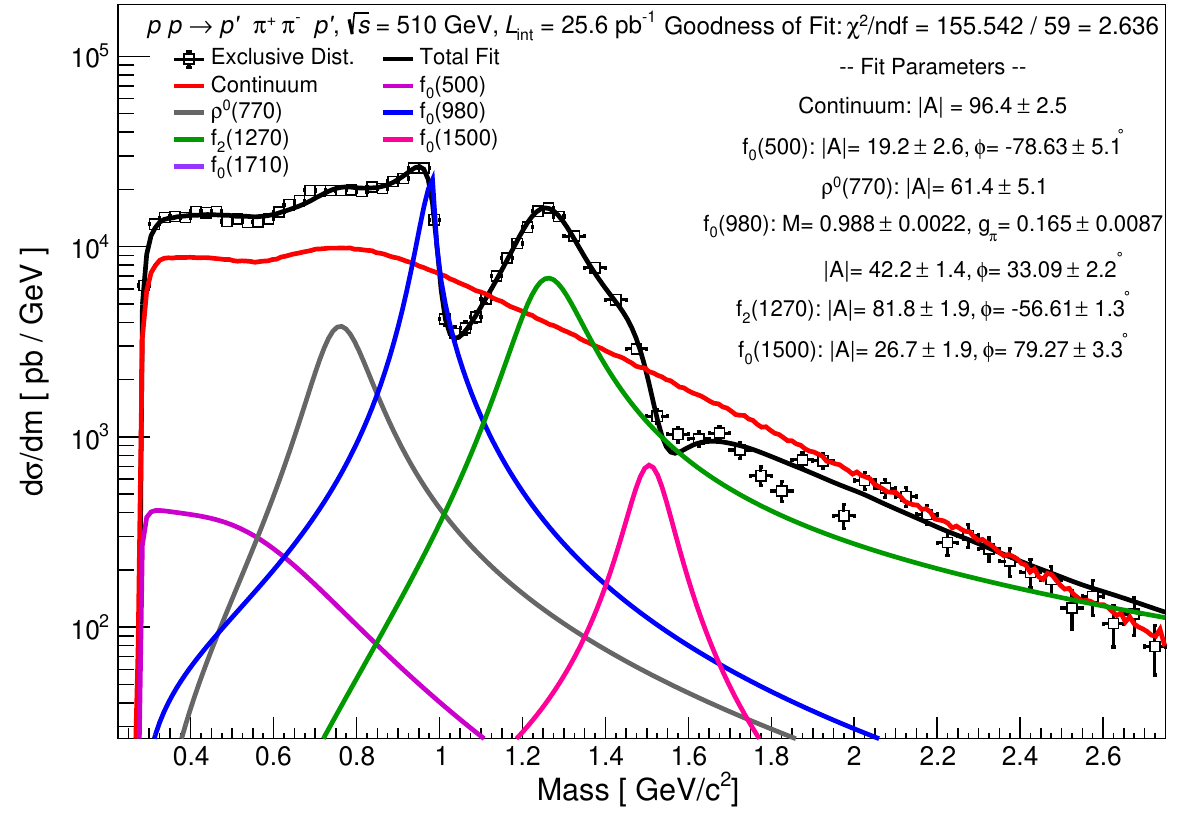}
 \caption{Background-subtracted $\pi^+\pi^-$ differential cross section with the baseline fit model and $f_0(1500)$.}
 \label{fig:pipi_baseline_1500}
\end{figure}

\begin{figure}[H]
 \includegraphics[width=0.45\textwidth]{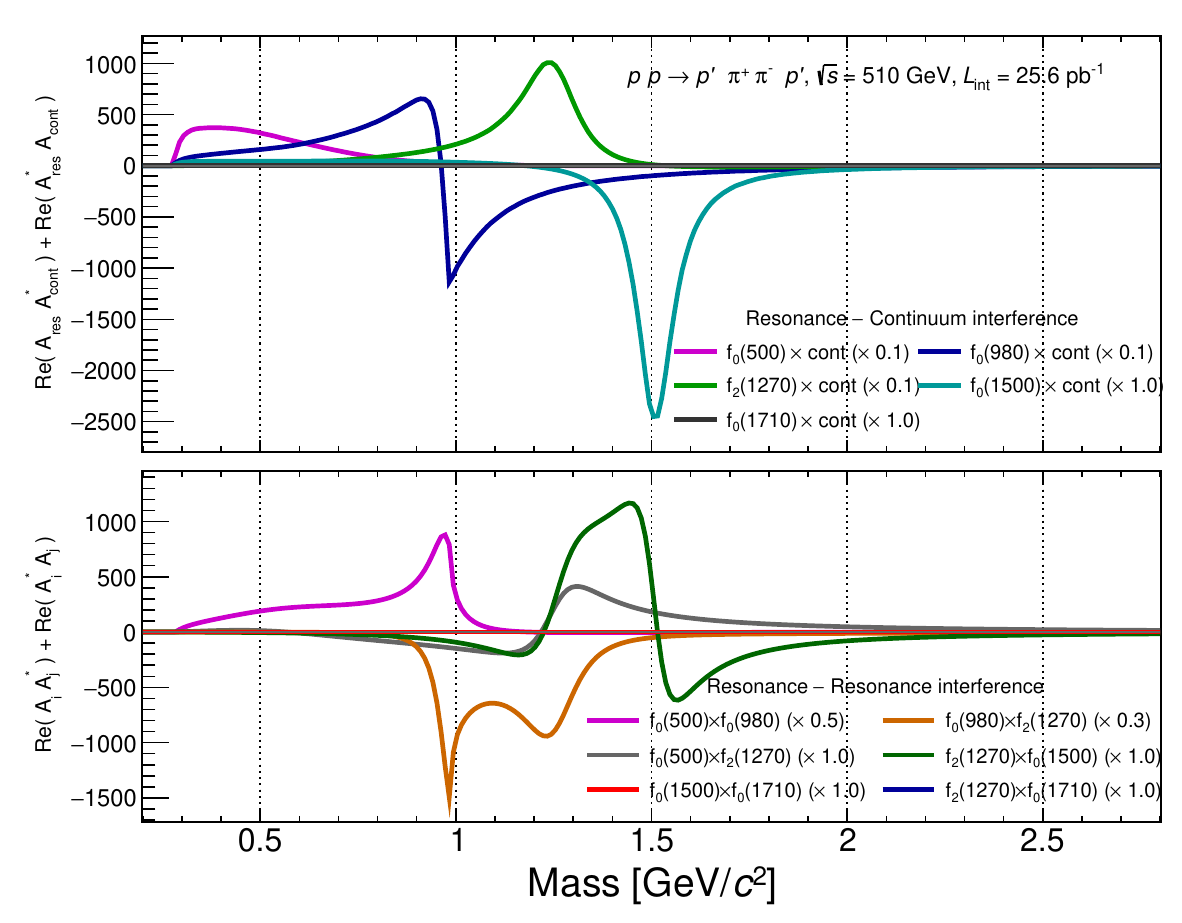}
 \caption{Interference contributions to the $\pi^+\pi^-$ differential cross section for figure~\ref{fig:pipi_baseline_1500}.}
 \label{fig:pipi_baseline_1500_interfernce}
\end{figure}

\begin{figure}[H]
 \includegraphics[width=0.45\textwidth]{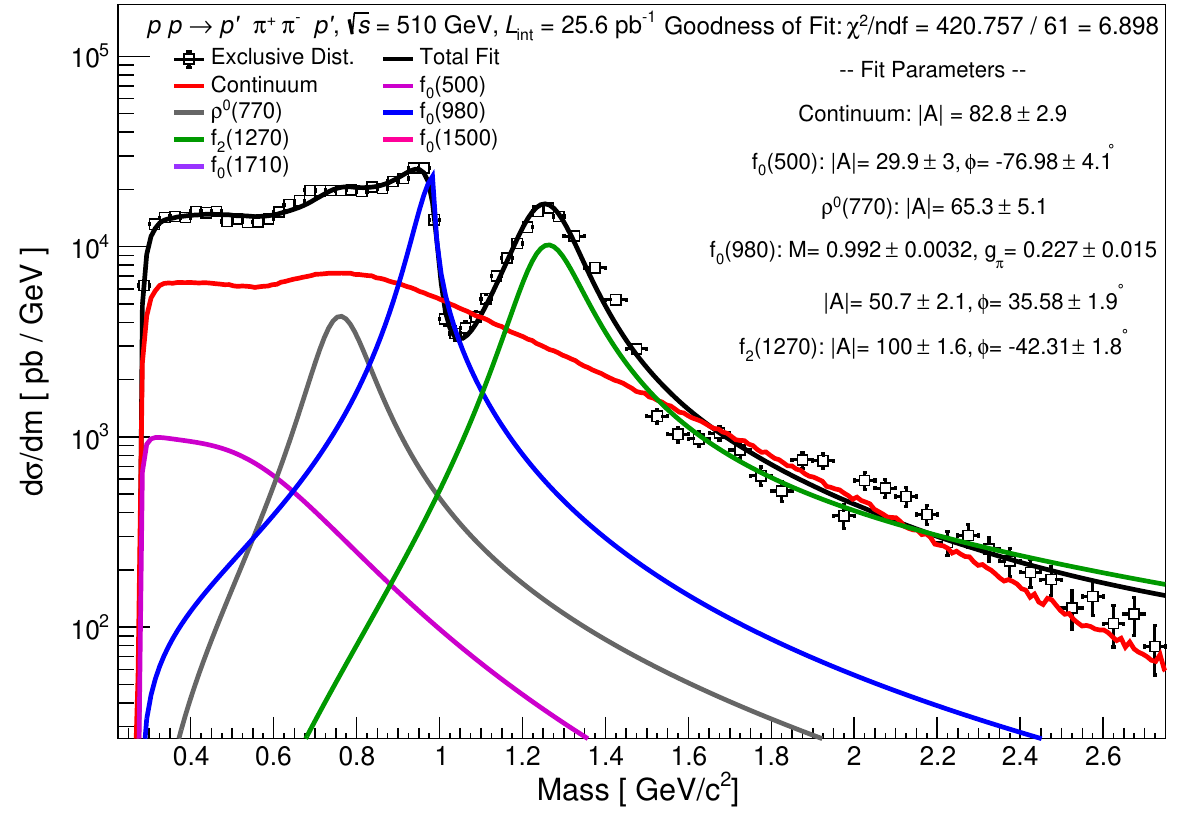}
 \caption{Background-subtracted $\pi^+\pi^-$ differential cross section with only baseline fit model.}
 \label{fig:pipi_baseline}
\end{figure}

\begin{figure}[H]
 \includegraphics[width=0.45\textwidth]{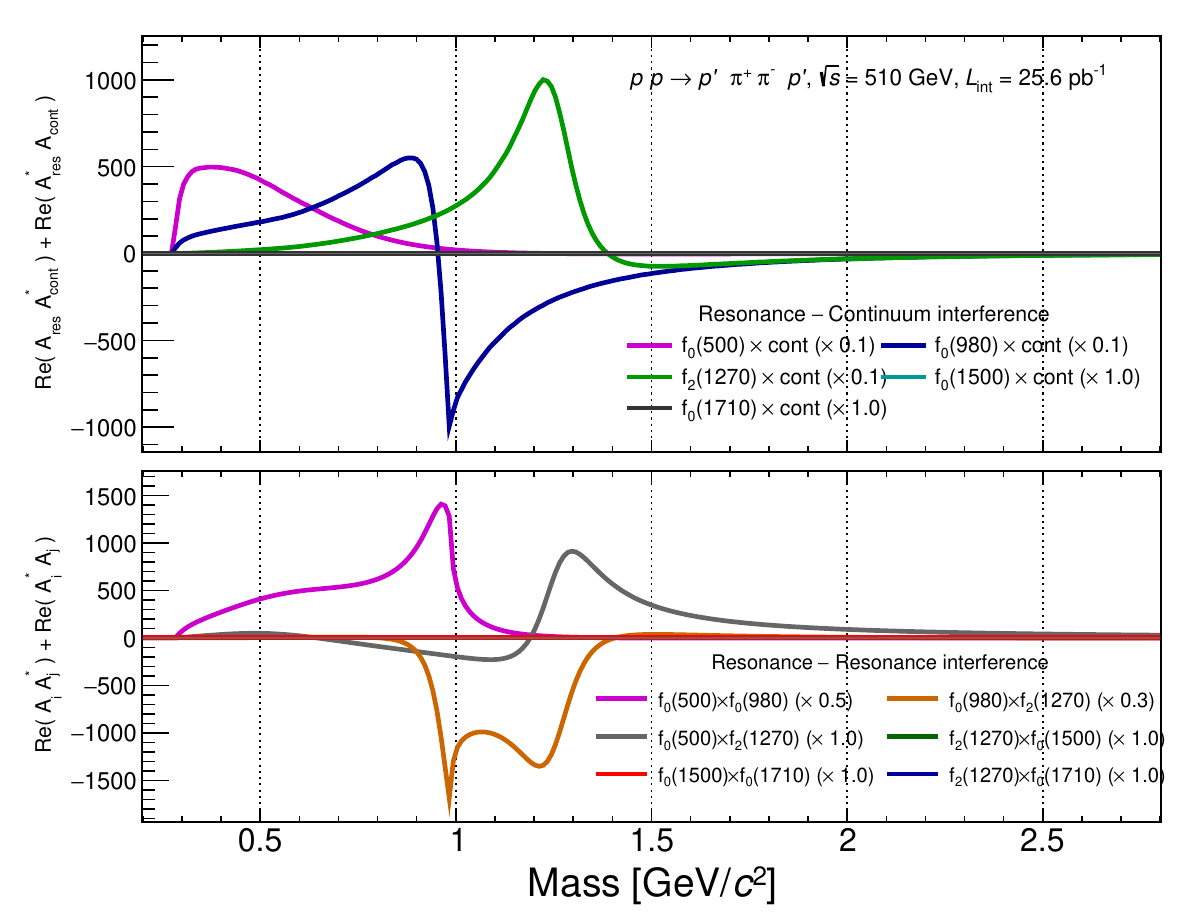}
 \caption{Interference contributions to the $\pi^+\pi^-$ differential cross section for figure~\ref{fig:pipi_baseline}.}
 \label{fig:pipi_baseline_interfernce}
\end{figure}

\begin{figure}[H]
 \includegraphics[width=0.45\textwidth]{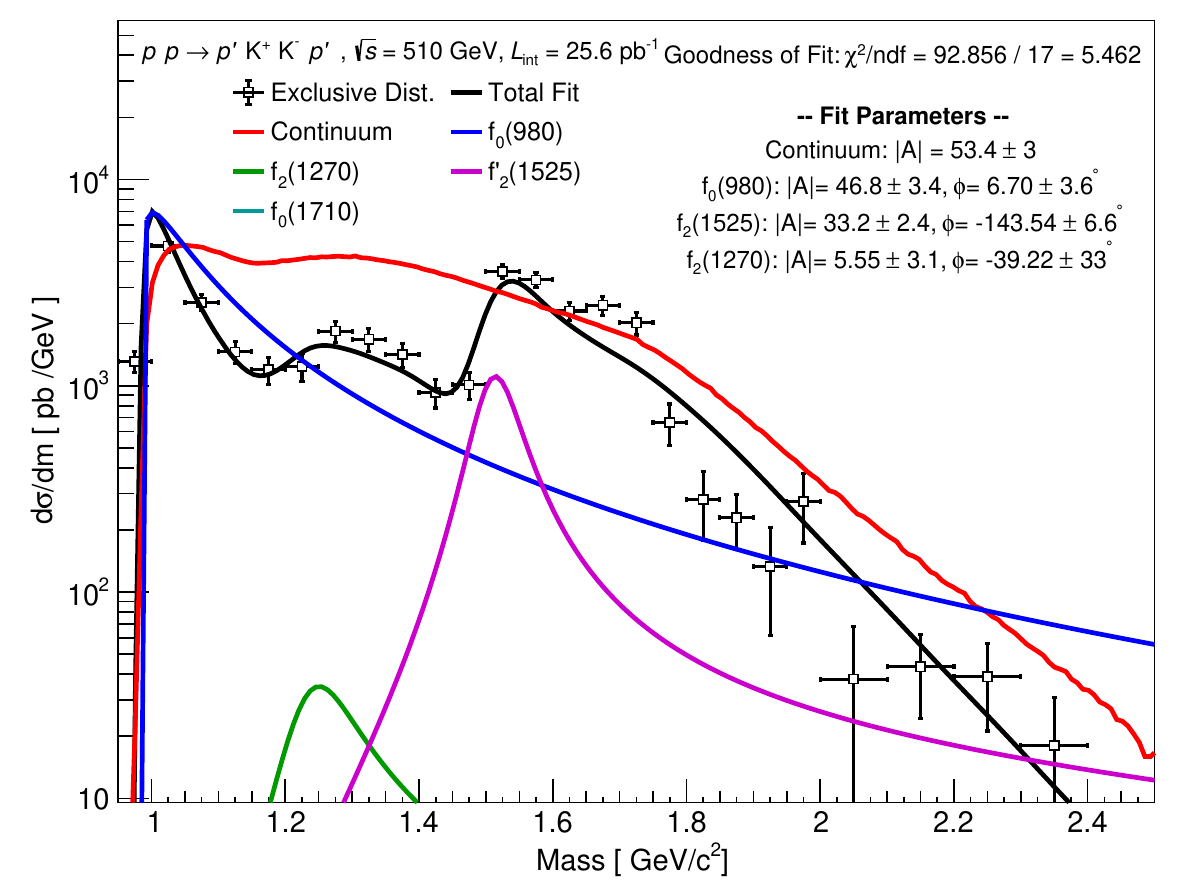}
 \caption{Background-subtracted $K^+K^-$ differential cross section with baseline fit model.}
 \label{fig:kk_baseline}
\end{figure}

\begin{figure}[H]
 
 \includegraphics[width=0.45\textwidth]{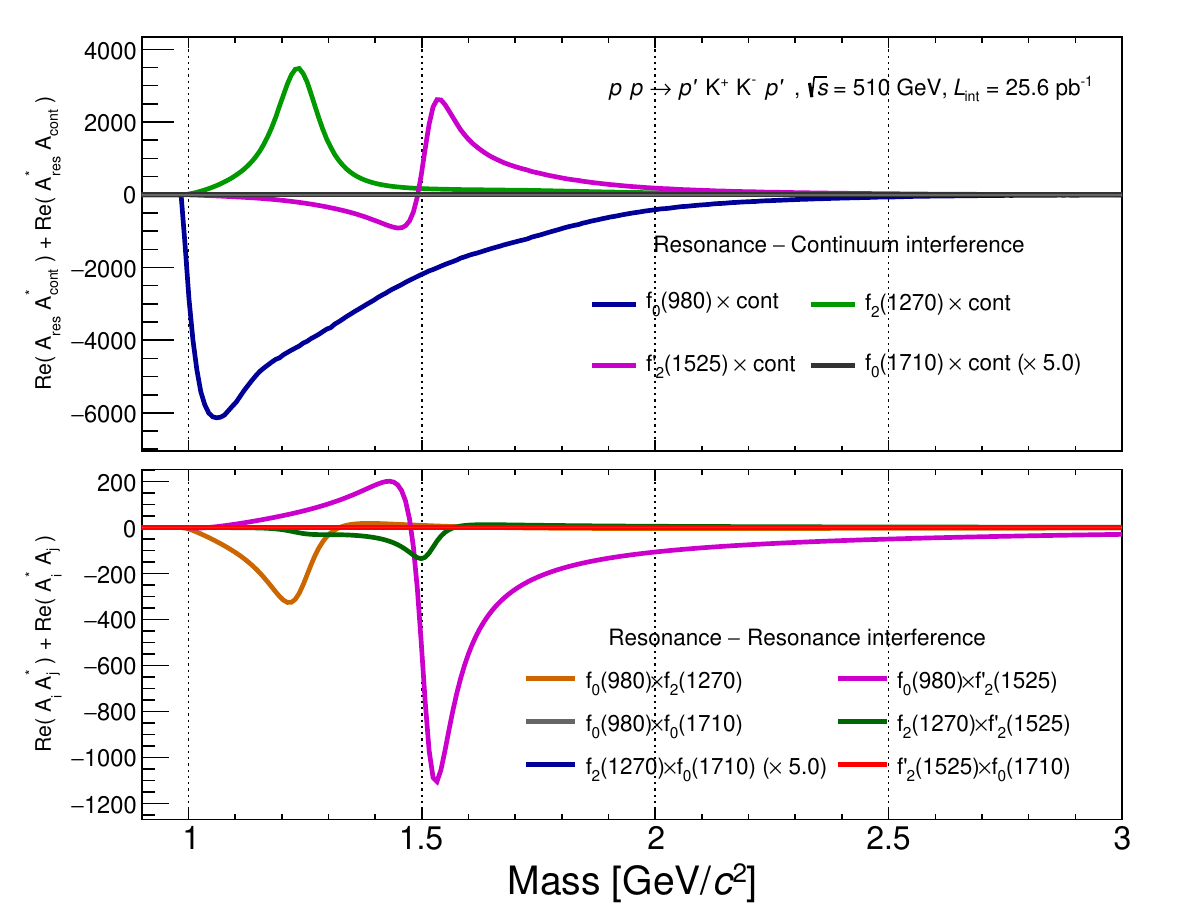}
 \caption{Interference contributions to the $K^+K^-$ differential cross section for figure~\ref{fig:kk_baseline}.}
 \label{fig:kk_baseline_interference}
\end{figure}


\section{Toy model for orthogonality breaking}
\label{app:toy_orthogonality}

An isotropic two-body decay generator is used to quantify the
nonorthogonality induced by fiducial acceptance in a real $S+D$
angular basis. No resonance model or physical production amplitude
is assumed. Natural units, $c=1$, are used, with the metric
$p^2=E^2-\mathbf{p}^2$. The construction proceeds as follows.

\begin{enumerate}

\item \textbf{Incoming beams.}

The West and East beams define the laboratory $+z$ and $-z$
directions, respectively. The beam momentum is
\begin{equation}
p_b = \sqrt{E_b^2-m_p^2},
\end{equation}
and the corresponding four-momenta are
\begin{equation}
\begin{aligned}
P_W^\mu &= (E_b,0,0,+p_b),\\
P_E^\mu &= (E_b,0,0,-p_b),
\end{aligned}
\end{equation}
where $E_b=255~\mathrm{GeV}$ and $m_p=0.938~\mathrm{GeV}$.

\item \textbf{Scattered protons.}

The distributions of $(\xi_E,\xi_W)$ and $(p_{T,E},p_{T,W})$ are obtained from events generated with the GRANIITTI Monte Carlo (MC) generator~\cite{Graniitti}, using the same collision configuration as the experimental data but without applying track selection criteria. Correlations
between the two sides are retained within each pair, while
joint $\xi$--$p_T$ correlations are not modeled. The proton
azimuthal angles are sampled independently according to
\begin{equation}
\phi_s \sim U(0,2\pi),
\qquad s\in\{W,E\}.
\end{equation}

The implemented momentum-loss definition and the components of
the outgoing proton four-momenta are
\begin{equation}
\begin{aligned}
|\mathbf{p}'_s|
    &= (1-\xi_s)p_b,\\
p'_{x,s}
    &= p_{T,s}\cos\phi_s,\\
p'_{y,s}
    &= p_{T,s}\sin\phi_s,\\
p'_{z,s}
    &= \sigma_s
       \sqrt{(1-\xi_s)^2p_b^2-p_{T,s}^2},\\
E'_s
    &= \sqrt{(1-\xi_s)^2p_b^2+m_p^2},\\
P_s'{}^\mu
    &= (E'_s,\mathbf{p}'_s),
\end{aligned}
\end{equation}
where $\sigma_W=+1$ and $\sigma_E=-1$. Physical configurations
satisfy
\begin{equation}
0\leq\xi_s<1,
\qquad
0\leq p_{T,s}\leq(1-\xi_s)p_b.
\end{equation}

\item \textbf{Central-system reconstruction.}

The central-system four-momentum is obtained from four-momentum
conservation. Defining the momentum transfers as
\begin{equation}
q_s^\mu=P_s^\mu-P_s'{}^\mu,
\end{equation}
the central-system kinematics are
\begin{equation}
\begin{aligned}
P_X^\mu
    &= q_W^\mu+q_E^\mu,\\
M
    &= \sqrt{P_X^2},\\
\boldsymbol{\beta}_X
    &= \frac{\mathbf{P}_X}{E_X}.
\end{aligned}
\end{equation}

Positive-energy, timelike configurations are retained subject to
\begin{equation}
M>2m_\pi,
\qquad
m_{\min}\leq M<m_{\max},
\end{equation}
where $m_\pi=0.13957~\mathrm{GeV}$. The default mass interval is
$[0.28,2.80)~\mathrm{GeV}$, divided into 63 bins of width
$0.04~\mathrm{GeV}$.

\item \textbf{Collins--Soper axes.}

The incoming beam four-vectors are boosted into the central-system
rest frame:
\begin{equation}
\begin{aligned}
P_s^{*\mu}
    &=\Lambda(-\boldsymbol{\beta}_X)P_s^\mu,\\
\hat{\mathbf{u}}_s
    &=\frac{\mathbf{P}_s^*}{|\mathbf{P}_s^*|}.
\end{aligned}
\end{equation}

Using the negative beam-sum convention, the Collins--Soper axes
are defined by
\begin{equation}
\begin{aligned}
\hat{\mathbf{z}}_{\mathrm{CS}}
&=
\frac{\hat{\mathbf{u}}_W-\hat{\mathbf{u}}_E}
     {|\hat{\mathbf{u}}_W-\hat{\mathbf{u}}_E|},
\\[2pt]
\mathbf{x}_0
&=
-\left(\hat{\mathbf{u}}_W+\hat{\mathbf{u}}_E\right),
\\[2pt]
\mathbf{x}_\perp
&=
\mathbf{x}_0
-
(\mathbf{x}_0\cdot\hat{\mathbf{z}}_{\mathrm{CS}})
\hat{\mathbf{z}}_{\mathrm{CS}},
\\[2pt]
\hat{\mathbf{x}}_{\mathrm{CS}}
&=
\frac{\mathbf{x}_\perp}{|\mathbf{x}_\perp|},
\\[2pt]
\hat{\mathbf{y}}_{\mathrm{CS}}
&=
\hat{\mathbf{z}}_{\mathrm{CS}}
\times
\hat{\mathbf{x}}_{\mathrm{CS}}.
\end{aligned}
\end{equation}

A final orthonormalization is applied numerically. If the
beam-sum transverse direction vanishes, a fixed laboratory
reference direction is projected perpendicular to
$\hat{\mathbf{z}}_{\mathrm{CS}}$ to define the azimuthal origin.

\item \textbf{Isotropic pion decay.}

The $\pi^+$ decay angles are generated uniformly in solid angle:
\begin{equation}
\begin{aligned}
c\equiv\cos\theta_{\mathrm{CS}}
    &\sim U(-1,1),\\
\phi_{\mathrm{CS}}
    &\sim U(0,2\pi),\\
\frac{dP}{d\Omega}
    &=\frac{1}{4\pi}.
\end{aligned}
\end{equation}

The pion direction in the central-system rest frame is
\begin{equation}
\begin{aligned}
\hat{\mathbf{n}}
={}&
\sqrt{1-c^2}\cos\phi_{\mathrm{CS}}\,
\hat{\mathbf{x}}_{\mathrm{CS}}
\\
&+
\sqrt{1-c^2}\sin\phi_{\mathrm{CS}}\,
\hat{\mathbf{y}}_{\mathrm{CS}}
+
c\,\hat{\mathbf{z}}_{\mathrm{CS}}.
\end{aligned}
\end{equation}

The pion momentum magnitude and four-vectors are then
\begin{equation}
\begin{aligned}
p_\pi^*
    &=\sqrt{\frac{M^2}{4}-m_\pi^2},\\
p_{\pi^\pm}^{*\mu}
    &=
    \left(
    \frac{M}{2},
    \pm p_\pi^*\hat{\mathbf{n}}
    \right),\\
p_{\pi^\pm}^{\mu,\mathrm{lab}}
    &=
    \Lambda(+\boldsymbol{\beta}_X)
    p_{\pi^\pm}^{*\mu}.
\end{aligned}
\end{equation}

By construction,
\begin{equation}
p_{\pi^+}^{\mu,\mathrm{lab}}
+
p_{\pi^-}^{\mu,\mathrm{lab}}
=
P_X^\mu.
\end{equation}

\item \textbf{Real angular basis.}

The ordered real angular basis is defined as
\begin{equation}
\mathbf{Y}
=
\left(
Y_{00},
Y_{20},
Y_{21c},
Y_{21s},
Y_{22c},
Y_{22s}
\right)^T,
\end{equation}
where $\phi=\phi_{\mathrm{CS}}$. The basis functions are
\begin{equation}
\begin{aligned}
Y_{00}
&=
\frac{1}{\sqrt{4\pi}},
\\
Y_{20}
&=
\sqrt{\frac{5}{16\pi}}\,(3c^2-1),
\\
Y_{21c}
&=
-\sqrt{\frac{15}{4\pi}}\,
c\sqrt{1-c^2}\cos\phi,
\\
Y_{21s}
&=
-\sqrt{\frac{15}{4\pi}}\,
c\sqrt{1-c^2}\sin\phi,
\\
Y_{22c}
&=
\sqrt{\frac{15}{16\pi}}\,
(1-c^2)\cos 2\phi,
\\
Y_{22s}
&=
\sqrt{\frac{15}{16\pi}}\,
(1-c^2)\sin 2\phi.
\end{aligned}
\end{equation}

These functions follow the Condon--Shortley convention and satisfy
the orthonormality condition
\begin{equation}
\int_{4\pi}
Y_a(\Omega)Y_b(\Omega)\,d\Omega
=
\delta_{ab}.
\end{equation}

\item \textbf{Successive acceptance stages.}

Three nested event samples are considered:
\begin{equation}
\begin{aligned}
\mathcal{S}_{\mathrm{MASS}}
&=
\{\text{events passing the mass selection}\},
\\
\mathcal{S}_{\mathrm{RP}}
&=
\mathcal{S}_{\mathrm{MASS}}
\cap
\{\text{both protons pass RP cuts}\},
\\
\mathcal{S}_{\mathrm{TPC}}
&=
\mathcal{S}_{\mathrm{RP}}
\cap
\{|\eta_{\pi^+}|\leq\eta_{\max},
  |\eta_{\pi^-}|\leq\eta_{\max}\}.
\end{aligned}
\end{equation}

The RP selection is based on the STAR Roman Pot acceptance at
$\sqrt{s}=510~\mathrm{GeV}$~\cite{STARCEP_510GeV}. The pion selection is applied using the
laboratory pseudorapidity,
\begin{equation}
\eta
=
-\ln\tan\left(\frac{\theta_{\mathrm{lab}}}{2}\right).
\end{equation}
\item \textbf{Mass-dependent acceptance matrices.}

For acceptance stage $s$ and mass bin $j$, the event count and
angular matrices are defined as
\begin{equation}
\begin{aligned}
N_{s,j}
&=
\#\{i\in\mathcal{S}_s:\,M_i\in j\},
\\
G_{ab}^{s,j}
&=
\sum_{\substack{i\in\mathcal{S}_s\\M_i\in j}}
Y_a(\Omega_i)Y_b(\Omega_i),
\\
K_{ab}^{s,j}
&=
\frac{G_{ab}^{s,j}}
{\sqrt{G_{aa}^{s,j}G_{bb}^{s,j}}}
\end{aligned}
\end{equation}
The matrix $G$ is the raw Gram matrix, $K$ quantifies the normalized
non-orthogonality of the angular basis after acceptance.

For complete angular coverage,
\begin{equation}
K\rightarrow I_6.
\end{equation}

Thus, nonuniform angular acceptance can prevent the $S$--$D_0$ interference term from vanishing upon angular integration. In the toy model, the MASS and RP selections are independent of the sampled pion decay angles, so their normalized overlap matrices approach $I_6$. In contrast, the restricted pion acceptance can introduce nonzero off-diagonal elements, indicating acceptance-induced non-orthogonality between the $S$- and $D_0$-wave components.

\end{enumerate}


\section{Correlation Matrices}
\label{app:correlation}

\begin{table*}[t]
\centering
\caption{Correlation matrix of free parameterss for full $\pi^+\pi^-$ fit}
\resizebox{\textwidth}{!}{%
\begin{tabular}{lrrrrrrrrrrrrrr}
\hline
 & $|A_{f_0(980)}|$ & $\phi_{f_0(980)}$ & $|A_{f_2(1270)}|$ & $\phi_{f_2(1270)}$ & $|A_{\rm cont}|$ & $|A_\rho|$ & $M_{f_0(980)}$ & $g_\pi$ & $|A_{f_0(500)}|$ & $\phi_{f_0(500)}$ & $|A_{f_0(1500)}|$ & $\phi_{f_0(1500)}$ & $|A_{f_0(1710)}|$ & $\phi_{f_0(1710)}$ \\
\hline
$|A_{f_0(980)}|$      &  1.000 &  0.737 &  0.240 & -0.038 &  0.474 & -0.562 &  0.705 &  0.721 & -0.544 &  0.259 & -0.171 & -0.200 & -0.119 &  0.017 \\
$\phi_{f_0(980)}$     &  0.737 &  1.000 & -0.223 & -0.180 &  0.729 & -0.544 &  0.801 &  0.266 & -0.722 &  0.194 &  0.105 & -0.357 &  0.112 & -0.111 \\
$|A_{f_2(1270)}|$     &  0.240 & -0.223 &  1.000 &  0.544 & -0.613 &  0.329 & -0.108 &  0.399 &  0.551 & -0.046 & -0.437 &  0.438 & -0.357 &  0.235 \\
$\phi_{f_2(1270)}$    & -0.038 & -0.180 &  0.544 &  1.000 & -0.481 &  0.295 & -0.177 &  0.017 &  0.461 & -0.105 & -0.460 &  0.032 & -0.527 &  0.065 \\
$|A_{\rm cont}|$      &  0.474 &  0.729 & -0.613 & -0.481 &  1.000 & -0.725 &  0.658 &  0.188 & -0.973 &  0.399 &  0.289 & -0.325 &  0.368 & -0.064 \\
$|A_\rho|$            & -0.562 & -0.544 &  0.329 &  0.295 & -0.725 &  1.000 & -0.569 & -0.429 &  0.718 &  0.004 & -0.099 &  0.263 & -0.161 &  0.034 \\
$M_{f_0(980)}$        &  0.705 &  0.801 & -0.108 & -0.177 &  0.658 & -0.569 &  1.000 &  0.523 & -0.677 &  0.201 &  0.013 & -0.271 &  0.062 & -0.037 \\
$g_\pi$               &  0.721 &  0.266 &  0.399 &  0.017 &  0.188 & -0.429 &  0.523 &  1.000 & -0.281 &  0.204 & -0.248 & -0.004 & -0.191 &  0.092 \\
$|A_{f_0(500)}|$      & -0.544 & -0.722 &  0.551 &  0.461 & -0.973 &  0.718 & -0.677 & -0.281 &  1.000 & -0.407 & -0.255 &  0.349 & -0.311 &  0.089 \\
$\phi_{f_0(500)}$     &  0.259 &  0.194 & -0.046 & -0.105 &  0.399 &  0.004 &  0.201 &  0.204 & -0.407 &  1.000 &  0.031 & -0.100 &  0.038 & -0.021 \\
$|A_{f_0(1500)}|$     & -0.171 &  0.105 & -0.437 & -0.460 &  0.289 & -0.099 &  0.013 & -0.248 & -0.255 &  0.031 &  1.000 &  0.109 &  0.243 & -0.468 \\
$\phi_{f_0(1500)}$    & -0.200 & -0.357 &  0.438 &  0.032 & -0.325 &  0.263 & -0.271 & -0.004 &  0.349 & -0.100 &  0.109 &  1.000 &  0.346 &  0.078 \\
$|A_{f_0(1710)}|$     & -0.119 &  0.112 & -0.357 & -0.527 &  0.368 & -0.161 &  0.062 & -0.191 & -0.311 &  0.038 &  0.243 &  0.346 &  1.000 &  0.095 \\
$\phi_{f_0(1710)}$    &  0.017 & -0.111 &  0.235 &  0.065 & -0.064 &  0.034 & -0.037 &  0.092 &  0.089 & -0.021 & -0.468 &  0.078 &  0.095 &  1.000 \\
\hline
\end{tabular}%
}
\label{tab:corr_matrix_pipi_baseline_1500_1710}
\end{table*}

\begin{table*}[]
\centering
\caption{Correlation matrix of free fit parameters for full $K^+K^-$ fit}
\label{tab:corr_matrix_kk_baseline_1710}
\scriptsize
\begin{tabular}{lrrrrrrrrr}
\hline
 & $|A_{f_0(980)}|$ & $\phi_{f_0(980)}$ & $|A_{f_2(1525)}|$ & $\phi_{f_2(1525)}$ & $|A_{\rm cont}|$ & $|A_{f_0(1710)}|$ & $\phi_{f_0(1710)}$ & $|A_{f_2(1270)}|$ & $\phi_{f_2(1270)}$ \\
\hline
$|A_{f_0(980)}|$      & 1.000 & 0.815 & 0.056 & 0.052 & 0.596 & -0.330 & -0.128 & -0.133 & 0.566 \\
$\phi_{f_0(980)}$     & 0.815 & 1.000 & 0.101 & -0.318 & 0.919 & -0.599 & -0.315 & -0.538 & 0.375 \\
$|A_{f_2(1525)}|$     & 0.056 & 0.101 & 1.000 & -0.378 & 0.133 & -0.173 & 0.303 & 0.041 & -0.035 \\
$\phi_{f_2(1525)}$    & 0.052 & -0.318 & -0.378 & 1.000 & -0.568 & 0.492 & 0.412 & 0.347 & 0.526 \\
$|A_{\rm cont}|$      & 0.596 & 0.919 & 0.133 & -0.568 & 1.000 & -0.737 & -0.408 & -0.701 & 0.072 \\
$|A_{f_0(1710)}|$     & -0.330 & -0.599 & -0.173 & 0.492 & -0.737 & 1.000 & 0.400 & 0.465 & 0.113 \\
$\phi_{f_0(1710)}$    & -0.128 & -0.315 & 0.303 & 0.412 & -0.408 & 0.400 & 1.000 & 0.408 & 0.197 \\
$|A_{f_2(1270)}|$     & -0.133 & -0.538 & 0.041 & 0.347 & -0.701 & 0.465 & 0.408 & 1.000 & -0.025 \\
$\phi_{f_2(1270)}$    & 0.566 & 0.375 & -0.035 & 0.526 & 0.072 & 0.113 & 0.197 & -0.025 & 1.000 \\
\hline
\end{tabular}
\end{table*}


\nocite{*}

\clearpage 

\bibliography{apssamp}

@misc{Vadacchino2023,
      title={A review on Glueball hunting}, 
      author={Davide Vadacchino},
      year={2023},
      eprint={2305.04869},
      archivePrefix={arXiv},
      primaryClass={hep-lat},
      url={https://arxiv.org/abs/2305.04869}, 
}

@article{CMS2020,
  author        = {Sirunyan, A. M. and Tumasyan, A. and Adam, W. and others},
  collaboration = { CMS Collaboration },
  title         = {Study of central exclusive $\pi^+\pi^-$ production in
                   proton-proton collisions at $\sqrt{s}=5.02$ and $13~\mathrm{TeV}$},
  journal       = {Eur. Phys. J. C},
  volume        = {80},
  pages         = {718},
  year          = {2020},
  doi           = {10.1140/epjc/s10052-020-8166-5},
  eprint        = {2003.02811},
  archivePrefix = {arXiv},
  primaryClass  = {hep-ex}
}

@article{WA102_pi+pi-,
  author        = {Barberis, D. and others},
  collaboration = { WA102 Collaboration },
  title         = {A partial wave analysis of the centrally produced
                   {$\pi^+\pi^-$} system in {$pp$} interactions at
                   {$450~\mathrm{GeV}/c$}},
  journal       = {Phys. Lett. B},
  volume        = {453},
  pages         = {316--324},
  year          = {1999},
  doi           = {10.1016/S0370-2693(99)00366-4},
  eprint        = {hep-ex/9903043},
  archivePrefix = {arXiv},
  primaryClass  = {hep-ex}
}

@article{WA102_KKbar,
  author        = {Barberis, D. and others},
  collaboration = { WA102 Collaboration },
  title         = {A partial wave analysis of the centrally produced
                   {$K^+K^-$} and {$K_S^0K_S^0$} systems in {$pp$}
                   interactions at {$450~\mathrm{GeV}/c$} and new information
                   on the spin of the {$f_J(1710)$}},
  journal       = {Phys. Lett. B},
  volume        = {453},
  pages         = {305--315},
  year          = {1999},
  doi           = {10.1016/S0370-2693(99)00365-2},
  eprint        = {hep-ex/9903042},
  archivePrefix = {arXiv},
  primaryClass  = {hep-ex}
}

@article{WA102_pi0pi0,
  author        = {Barberis, D. and others},
  collaboration = { WA102 Collaboration },
  title         = {A partial wave analysis of the centrally produced
                   {$\pi^0\pi^0$} system in {$pp$} interactions at
                   {$450~\mathrm{GeV}/c$}},
  journal       = {Phys. Lett. B},
  volume        = {453},
  pages         = {325--332},
  year          = {1999},
  doi           = {10.1016/S0370-2693(99)00367-6},
  eprint        = {hep-ex/9903044},
  archivePrefix = {arXiv},
  primaryClass  = {hep-ex}
}

@article{WA102_4pion,
  author        = {Barberis, D. and others},
  collaboration = { WA102 Collaboration },
  title         = {A study of the centrally produced
                   $\pi^+\pi^-\pi^+\pi^-$ channel in $pp$ interactions
                   at $450~\mathrm{GeV}/c$},
  journal       = {Phys. Lett. B},
  volume        = {413},
  pages         = {217--224},
  year          = {1997},
  doi           = {10.1016/S0370-2693(97)01053-1},
  eprint        = {hep-ex/9707021},
  archivePrefix = {arXiv}
}

@article{WA102_EtaEta,
  author        = {Barberis, D. and others},
  collaboration = { WA102 Collaboration },
  title         = {A study of the {$\eta\eta$} channel produced in central
                   {$pp$} interactions at {$450~\mathrm{GeV}/c$}},
  journal       = {Phys. Lett. B},
  volume        = {479},
  pages         = {59--66},
  year          = {2000},
  doi           = {10.1016/S0370-2693(00)00374-9},
  eprint        = {hep-ex/0003033},
  archivePrefix = {arXiv},
  primaryClass  = {hep-ex}
}

@article{WA102_coupled_channel,
  author        = {Barberis, D. and others},
  collaboration = { WA102 Collaboration },
  title         = {A coupled channel analysis of the centrally produced
                   {$K^{+}K^{-}$} and {$\pi^{+}\pi^{-}$} final states in
                   {$pp$} interactions at {$450~\mathrm{GeV}/c$}},
  journal       = {Phys. Lett. B},
  volume        = {462},
  pages         = {462--470},
  year          = {1999},
  doi           = {10.1016/S0370-2693(99)00909-0},
  eprint        = {hep-ex/9907055},
  archivePrefix = {arXiv},
  primaryClass  = {hep-ex}
}

@article{CloseKirk1997,
  author  = {F. E. Close and A. Kirk},
  title   = {Glueball-$q\bar{q}$ filter in central hadron production},
  journal = {Phys. Lett. B},
  volume  = {397},
  pages   = {333--338},
  year    = {1997},
  doi     = {10.1016/S0370-2693(97)00222-0}
}

@article{CloseKirk2001,
  author  = {F. E. Close and A. Kirk},
  title   = {Scalar Glueball--$q\bar q$ Mixing above 1 GeV and Implications for Lattice QCD},
  journal = {European Physical Journal C},
  volume  = {21},
  pages   = {531--543},
  year    = {2001},
  doi     = {10.1007/s100520100748}
}

@article{STARCEP_510GeV,
  author        = {Aboona, B. E. and others},
  collaboration = {STAR Collaboration},
  title         = {Study of Central Exclusive Production of
                   {$\pi^+\pi^-$}, {$K^+K^-$} and {$p\bar{p}$} pairs
                   in proton-proton collisions at {$\sqrt{s}=510~\mathrm{GeV}$}
                   with the {STAR} detector at {RHIC}},
  journal       = {JHEP},
  volume        = {07},
  number        = {143},
  pages         = {143},
  year          = {2026},
  doi           = {10.1007/JHEP07(2026)143},
  eprint        = {2510.27482},
  archivePrefix = {arXiv},
  primaryClass  = {hep-ex}
}

@article{STARDetector,
  author        = {Adam, Jaroslav and Cherney, Michael G. and D'Alesio, Joey
                   and Dufresne, Emma and Holub, Lukáš and Seger, Janet E.
                   and Tlustý, David},
  title         = {Upgrade of hardware controls for the STAR experiment at RHIC},
  journal       = {Nucl. Instrum. Meth. A},
  volume        = {1013},
  pages         = {165644},
  year          = {2021},
  doi           = {10.1016/j.nima.2021.165644},
  eprint        = {2103.09231},
  archivePrefix = {arXiv},
  primaryClass  = {physics.ins-det}
}

@article{Graniitti,
  author        = {Mieskolainen, Mikael},
  title         = {{GRANIITTI}: A Monte Carlo Event Generator for High Energy Diffraction},
  journal       = {arXiv:1910.06300},
  eprint        = {1910.06300},
  archivePrefix = {arXiv},
  primaryClass  = {hep-ph},
  year          = {2019}
}

@misc{Graniitti_FitTo200GeV,
  author        = {Mieskolainen, Mikael},
  title         = {{GRANIITTI}: Towards a Deep Learning-Enhanced Monte
                   Carlo Event Generator for High-Energy Diffraction},
  year          = {2023},
  eprint        = {2304.06010},
  archivePrefix = {arXiv},
  primaryClass  = {hep-ph}
}

@article{GarciaMartin2011,
  author        = {García-Martín, R. and Kamiński, R. and Peláez, J. R.
                   and Ruiz de Elvira, J.},
  title         = {Precise Determination of the {$f_0(600)$} and {$f_0(980)$}
                   Pole Parameters from a Dispersive Data Analysis},
  journal       = {Phys. Rev. Lett.},
  volume        = {107},
  pages         = {072001},
  year          = {2011},
  doi           = {10.1103/PhysRevLett.107.072001},
  eprint        = {1107.1635},
  archivePrefix = {arXiv},
  primaryClass  = {hep-ph}
}

@article{PDG2024,
  author        = {Navas, S. and others},
  collaboration = {Particle Data Group},
  title         = {Review of Particle Physics},
  journal       = {Phys. Rev. D},
  volume        = {110},
  pages         = {030001},
  year          = {2024},
  doi           = {10.1103/PhysRevD.110.030001}
}

@article{STAR_CEP_200GeV,
  author        = {Adam, J. and others},
  collaboration = {STAR Collaboration},
  title         = {Measurement of the Central Exclusive Production of
                   Charged Particle Pairs in Proton-Proton Collisions at
                   {$\sqrt{s}=200~\mathrm{GeV}$} with the {STAR} Detector
                   at {RHIC}},
  journal       = {JHEP},
  volume        = {07},
  number        = {178},
  pages         = {178},
  year          = {2020},
  doi           = {10.1007/JHEP07(2020)178},
  eprint        = {2004.11078},
  archivePrefix = {arXiv},
  primaryClass  = {hep-ex}
}

@article{Wilks_theorem,
  author  = {Algeri, Sara and Aalbers, Jelle and Mor{\aa}, Knut Dundas and Conrad, Jan},
  title   = {Searching for new phenomena with profile likelihood ratio tests},
  journal = {Nature Reviews Physics},
  volume  = {2},
  pages   = {245--252},
  year    = {2020},
  doi     = {10.1038/s42254-020-0169-5}
}

@article{PDG2026,
    author        = {Takahashi, F. and others},
    collaboration = {Particle Data Group},
    title         = {Review of Particle Physics},
    journal       = {Int. J. Mod. Phys. A},
    volume        = {41},
    pages         = {2630011},
    year          = {2026}
}

@article{radiative_jpsi_pipi,
  author        = {Ablikim, M. and others},
  collaboration = {BES Collaboration},
  title         = {Partial wave analyses of {$J/\psi \to \gamma\pi^+\pi^-$}
                   and {$\gamma\pi^0\pi^0$}},
  journal       = {Phys. Lett. B},
  volume        = {642},
  pages         = {441--448},
  year          = {2006},
  doi           = {10.1016/j.physletb.2006.10.004},
  eprint        = {hep-ex/0603048},
  archivePrefix = {arXiv},
  primaryClass  = {hep-ex}
}

@article{radiative_jpsi_kk,
  author        = {Bai, J. Z. and others},
  collaboration = {BES},
  title         = {Partial wave analyses of {$J/\psi \to \gamma K^+K^-$}
                   and {$\gamma K_S^0 K_S^0$}},
  journal       = {Phys. Rev. D},
  volume        = {68},
  pages         = {052003},
  year          = {2003},
  doi           = {10.1103/PhysRevD.68.052003},
  eprint        = {hep-ex/0307058},
  archivePrefix = {arXiv},
  primaryClass  = {hep-ex}
}

@article{radiative_jpsi_mixing,
  author        = {Close, Frank E. and Zhao, Qiang},
  title         = {Production of {$f_0(1710)$}, {$f_0(1500)$}, and
                   {$f_0(1370)$} in {$J/\psi$} hadronic decays},
  journal       = {Phys. Rev. D},
  volume        = {71},
  pages         = {094022},
  year          = {2005},
  doi           = {10.1103/PhysRevD.71.094022},
  eprint        = {hep-ph/0504043},
  archivePrefix = {arXiv},
  primaryClass  = {hep-ph}
}

@article{ChengChuaLiu2006,
  author        = {Cheng, Hai-Yang and Chua, Chun-Khiang and Liu, Keh-Fei},
  title         = {Scalar glueball, scalar quarkonia, and their mixing},
  journal       = {Phys. Rev. D},
  volume        = {74},
  pages         = {094005},
  year          = {2006},
  doi           = {10.1103/PhysRevD.74.094005},
  eprint        = {hep-ph/0607206},
  archivePrefix = {arXiv},
  primaryClass  = {hep-ph}
}

@article{JanowskiGiacosaRischke2014,
  author        = {Janowski, Stanislaus and Giacosa, Francesco and Rischke, Dirk H.},
  title         = {Is {$f_0(1710)$} a glueball?},
  journal       = {Phys. Rev. D},
  volume        = {90},
  pages         = {114005},
  year          = {2014},
  doi           = {10.1103/PhysRevD.90.114005},
  eprint        = {1408.4921},
  archivePrefix = {arXiv},
  primaryClass  = {hep-ph}
}

@article{ALEPH_gg_glueball,
  author        = {Barate, R. and others},
  collaboration = {ALEPH},
  title         = {Search for the glueball candidates {$f_0(1500)$} and
                   {$f_J(1710)$} in {$\gamma\gamma$} collisions},
  journal       = {Phys. Lett. B},
  volume        = {472},
  pages         = {189--199},
  year          = {2000},
  doi           = {10.1016/S0370-2693(99)01390-8},
  eprint        = {hep-ex/9911022},
  archivePrefix = {arXiv},
  primaryClass  = {hep-ex}
}

@article{QuenchedLatticeGlueball_Old,
  author        = {Bali, G. S. and Schilling, K. and Hulsebos, A. and
                   Irving, A. C. and Michael, C. and Stephenson, P. W.},
  collaboration = {UKQCD Collaboration},
  title         = {A Comprehensive Lattice Study of {$SU(3)$} Glueballs},
  journal       = {Phys. Lett. B},
  volume        = {309},
  pages         = {378--384},
  year          = {1993},
  doi           = {10.1016/0370-2693(93)90948-H},
  eprint        = {hep-lat/9304012},
  archivePrefix = {arXiv},
  primaryClass  = {hep-lat}
}

@article{QuenchedLatticeGlueball_New,
  author        = {Chen, Y. and Alexandru, A. and Dong, S. J. and
                   Draper, T. and Horvath, I. and Lee, F. X. and
                   Liu, K. F. and Mathur, N. and Morningstar, C. and
                   Peardon, M. and Tamhankar, S. and Young, B. L. and
                   Zhang, J. B.},
  title         = {Glueball spectrum and matrix elements on anisotropic lattices},
  journal       = {Phys. Rev. D},
  volume        = {73},
  pages         = {014516},
  year          = {2006},
  doi           = {10.1103/PhysRevD.73.014516},
  eprint        = {hep-lat/0510074},
  archivePrefix = {arXiv},
  primaryClass  = {hep-lat}
}

@article{Belle_KSKS_2013,
  author        = {Uehara, S. and others},
  collaboration = {Belle Collaboration},
  title         = {High-statistics study of {$K_S^0 K_S^0$} pair production
                   in two-photon collisions},
  journal       = {Prog. Theor. Exp. Phys.},
  volume        = {2013},
  pages         = {123C01},
  year          = {2013},
  doi           = {10.1093/ptep/ptt097},
  eprint        = {1307.7457},
  archivePrefix = {arXiv},
  primaryClass  = {hep-ex}
}

@article{Belle_pi0pi0,
  author        = {Uehara, S. and others},
  collaboration = {Belle Collaboration},
  title         = {High-statistics study of neutral-pion pair production
                   in two-photon collisions},
  journal       = {Phys. Rev. D},
  volume        = {79},
  pages         = {052009},
  year          = {2009},
  doi           = {10.1103/PhysRevD.79.052009},
  eprint        = {0903.3697},
  archivePrefix = {arXiv},
  primaryClass  = {hep-ex}
}

@article{JinShen2021,
  author  = {Jin, Shan and Shen, Xiaoyan},
  title   = {Highlights of light meson spectroscopy at the {BESIII} experiment},
  journal = {Natl. Sci. Rev.},
  volume  = {8},
  number  = {11},
  pages   = {nwab198},
  year    = {2021},
  doi     = {10.1093/nsr/nwab198}
}

@article{WA102_CEP_mixing,
  author        = {Close, Frank E. and Kirk, Andrew},
  title         = {The mixing of {$f_0(1370)$}, {$f_0(1500)$} and
                   {$f_0(1710)$} and the search for the scalar glueball},
  journal       = {Phys. Lett. B},
  volume        = {483},
  pages         = {345--352},
  year          = {2000},
  doi           = {10.1016/S0370-2693(00)00623-7},
  eprint        = {hep-ph/0004241},
  archivePrefix = {arXiv},
  primaryClass  = {hep-ph}
}

@article{Kirk_OmegaSpectrometer_Review,
  author        = {Kirk, A.},
  title         = {Resonance production in central {$pp$} collisions at the
                   {CERN} {Omega} Spectrometer},
  journal       = {Phys. Lett. B},
  volume        = {489},
  pages         = {29--37},
  year          = {2000},
  doi           = {10.1016/S0370-2693(00)00944-8},
  eprint        = {hep-ph/0008053},
  archivePrefix = {arXiv},
  primaryClass  = {hep-ph}
}

@article{Ablikim2005Resonances,
  author  = {Ablikim, M. and others},
  collaboration = {Belle Collaboration},
  title   = {Resonances in $J/\psi \to \phi\pi^+\pi^-$ and
             $\phi K^+K^-$},
  journal = {Physics Letters B},
  volume  = {607},
  number  = {3-4},
  pages   = {243--253},
  year    = {2005},
  issn    = {0370-2693},
  doi     = {10.1016/j.physletb.2004.12.041},
  url     = {https://doi.org/10.1016/j.physletb.2004.12.041}
}

\end{document}